%% file: JLT2020_final.tex
\documentclass[journal]{IEEEtran}
\usepackage{amsmath,amssymb,amscd,latexsym,dsfont}
\usepackage{comment}
\usepackage{color}
\usepackage{subfigure}
\usepackage{enumerate}
\usepackage{enumitem}
\usepackage{stfloats}
\usepackage{psfrag}
\usepackage{setspace}
\usepackage{cite}
\usepackage{balance}
\usepackage{enumitem}
\usepackage{tikz}
\usepackage{pgfplots,amsfonts}
\usepackage{pgfplotstable}
\usetikzlibrary{shapes}
\usetikzlibrary{spy}
\usetikzlibrary{fit}
\usetikzlibrary{shapes.multipart}
\usetikzlibrary{positioning}
\input{myFigures.tex}

\input{myStyles.tex}
\input{myMacros.tex}

\input{myPackages.tex}
\pgfplotsset{compat=1.14}
\usepackage{wrapfig}
\usepackage[acronym,nomain]{glossaries}
\input{acronyms.tex}

\usepackage[ruled,linesnumbered]{algorithm2e}

\usepackage{arydshln}

\usepackage{bbm} 

\IEEEoverridecommandlockouts

\title{Analysis and Experimental Demonstration of Orthant-Symmetric Four-dimensional 7~bit/4D-sym Modulation for Optical Fiber Communication}

\author{Bin~Chen,~\IEEEmembership{Member,~IEEE},
Alex~Alvarado,~\IEEEmembership{Senior Member,~IEEE},\\ Sjoerd~van~der~Heide,~\IEEEmembership{Student Member,~IEEE}, Menno~van~den~Hout,~\IEEEmembership{Student Member,~IEEE},\\ Hartmut~Hafermann, \IEEEmembership{Senior Member,~IEEE},  and  Chigo~Okonkwo,~\IEEEmembership{Senior Member,~IEEE}.
\thanks{B. Chen is with  the School of Computer Science and Information Engineering, Hefei University of Technology, Hefei, China.
B. Chen is also with Eindhoven University of Technology, 5600 MB, Eindhoven, The Netherlands  (e-mail:~bin.chen@hfut.edu.cn).}
\thanks{A. Alvarado is with the Information and Communication Theory Lab, Signal Processing Systems Group, Department of Electrical Engineering, Eindhoven University of Technology, 5600 MB, Eindhoven, The Netherlands (e-mail:~a.alvarado@tue.nl).}
\thanks{S. van der Heide, M. van den Hout and C. Okonkwo  are with the High Capacity Optical Transmission Laboratory, Eindhoven University of Technology, Eindhoven 5600 MB, The Netherlands (e-mail:~\{s.p.v.d.heide, m.v.d.hout, c.m.okonkwo\}@tue.nl).}
\thanks{H. Hafermann is with the Optical Communication Technology Lab, Paris Research Center, Huawei Technologies France SASU, 92100 Boulogne-Billancourt, France (e-mail:~hartmut.hafermann@huawei.com).}
\thanks{
This research is supported in part by Huawei
France through the NLCAP project. 
B. Chen is supported  by the National Natural Science Foundation
of China (NSFC) under Grant 61701155, Fundamental Research Funds for the Central Universities (JZ2020HGTB0015),  and Anhui Provincial Natural Science Foundation (2008085QF282).  S. van der Heide, M. van den Hout and C. Okonkwo are supported in part by the Dutch NWO Gravitation Program: Research centre for Integrated Nanophotonics (Grant Number
024.002.033).   A. Alvarado is supported by the Netherlands Organisation for Scientific Research (NWO) via the VIDI Grant ICONIC (project number 15685). }
}

\begin{document}
\maketitle

\begin{abstract}
We propose a new four-dimensional  orthant-symmetric 128-ary modulation format (4D-OS128) with a  spectral efficiency of 7~bit/4D-sym. The proposed format fills the gap between polarization-multiplexed 8- and 16-ary quadrature-amplitude modulation (PM-8QAM and PM-16QAM).
Numerical simulations show that 4D-OS128 outperforms  two well-studied 4D geometrically-shaped modulation formats: 128SP-16QAM and 7b4D-2A8PSK by up to 0.65~dB  for  bit-interleaved  coded  modulation  at  the same spectral efficiency.
These gains are experimentally demonstrated in a  11$\times$233~Gbit/s wavelength division multiplexing (WDM) transmission system operating at 5.95 bit/4D-sym over 6000~km and 9000~km for both EDFA-only and hybrid amplification scenarios, respectively. A reach increase of 15\% is achieved with respect to 128-ary set-partitioning 16QAM.
Furthermore, the proposed 4D-OS128 is also compared to  $\boldsymbol{D}_4$ lattice-based constellation, 16QAM and probabilistically-shaped 16QAM with finite blocklength  via simulation.
\end{abstract}

\vspace{-0.5em}
\section{Introduction}

The demand for higher capacity and longer transmission distances in  optical fiber communications has been growing for several years.
In order  to  further  support  the  exponential  traffic growth, various multiplexing or modulation dimensions such as time, polarization,   wavelength   and  space   (multi-mode/multi-core fibers),  have  been  used.  In particular, \gls{WDM}  transport systems with coherent detection  have been   studied to improve \gls{SE} through high-order  modulation formats.
These formats have been shown to be promising for  high-speed long-haul transmission systems \cite{EssiambreJLT2010}. 

Achieving higher \gls{SE} by employing  polarization multiplexed $M$-ary quadrature  amplitude  modulation  (PM-$M$QAM) formats  {with the same \gls{FEC} overhead (OH)}  comes at the cost of a reduced transmission reach and has been extensively studied \cite{Bulow:09,Winzer:10}. 
In order to maintain or extend transmission distances in high-speed fiber-optic systems, \emph{signal shaping} has recently been the focus of considerable attention in the optical communications community. 

Shaping methods can be broadly categorized into \gls{PS} and \gls{GS}, both having distinct advantages and disadvantages. In \gls{PS}, long coded sequences induce  nonuniform probability distribution on the constellation points \cite{Buchali2016,TobiasJLT16,BochererECOC2017,Buchali2017,Maher2017,Amari2019_IntroducingESSoptics,Goossens2019_FirstExperimentESS}.
\gls{GS} employs a uniform distribution (i.e., equiprobable symbols) on non-equidistant constellation points \cite{Qu2017,ZhangECOC2017,BinECOC2018,BinICTON2018}.
Despite the difference between  GS and PS schemes, both techniques are employing the dimensionality to induce a nonuniform distribution via  either coding or  multidimensional signal space.
For  the  additive  white  Gaussian  noise (AWGN) channel, it has been shown that \gls{PS} outperforms \gls{GS} and more closely approaches Shannon’s channel capacity  when  the  number  of  constellation  points  is limited \cite[Sec.~4.3]{Szczecinski2015BICM},\cite{SteinerSCC2017}. On the other hand, \gls{GS} over multiple dimensions can not only reduce the gap to the Shannon capacity \cite{AgrellJLT2009,Karlsson:09,KoikeAkinoECOC2013,Millar:14,Millar2018_OFC}, but could also mitigate the nonlinear effects in the optical channel  \cite{Chagnon:13,Shiner:14,ReimerOFC2016,Kojima2017JLT,BinChenJLT2019, BinChenPTL2019}.  {Multidimensional (MD) \gls{GS}
 relies only on the selection of the location of constellation points in a (relatively low) MD space and the design of the corresponding MD detector. MD-GS therefore offers an interesting approach to achieve  shaping gains with low implementation complexity instead of using  coding approach. 
 MD-GS   can also be easily coupled with FEC and only requires straightforward modifications of the mapper and demapper.
However, MD-GS also increases the computational complexity of the demapper, as in this case, Euclidean distances for all multidimensional
symbols need to be calculated.  Nevertheless,  it has been shown \cite{YoshidaECOC2016,BendimeradECOC2018,NakamuraJLT2018} that low-complexity MD soft demapper schemes can be designed to achieve good trade-offs
between performance and complexity.}
 
Four-dimensional (4D) modulation formats are typically optimized in the four dimensions consisting of the two quadratures (I/Q) and the two polarization states (X/Y) of the optical field. These formats are often designed to achieve large minimum Euclidean distances \cite{AgrellJLT2009,Karlsson:09,KoikeAkinoECOC2013,Millar:14}. Conventional polarization multiplexed formats are not \emph{true} 4D formats, because they are only optimized in I, Q, X, and Y independently.   {4D  modulation  formats with dependency between dimensions can be obtained by  using sphere packing arguments\footnote{An excellent summary of MD constellations is given in the online database \cite{ErikDatabase}.}, applying Ungerboeck’s set-partitioning (SP) scheme \cite{UngerboeckTIT1982} or warping all the points to be nonuniformly spaced coordinates.}
{For most of good spherically bounded constellations, Gray labeling  does not exist, and therefore, the combination with binary FEC and \gls{BICM} results in a loss in \gls{AIRs} \cite{WaeckerleSCC2015,Alvarado2015_JLT}.}
Set-partitioning (SP) PM-16QAM has been investigated to  achieve fine granularity as  32-ary set-partitioning QAM (32SP-16QAM) \cite{MullerECOC2013,SunECOC2013}, 64-ary set-partitioning QAM (64SP-16QAM) \cite{NakamuraECOC2015} and 128-ary set-partitioning QAM (128SP-16QAM) \cite{ErikssonOE13,KashiECOC2015,KaihuiOFT2018}. 32SP-QAM, 64SP-QAM, and 128SP-QAM have 5, 6, 7 bit/4D-sym, resp., and can be used to achieve different transmission distances.
4D set-partitioned modulation formats based on the  $\boldsymbol{D}_4$ lattice have also been investigated by using multilevel coding (MLC) and multi-stage decoding (MSD) in fiber-optical communications \cite{FreyECOC2019,FreyJLT2020}.
The \textit{warping} technique uses constellation points from a nonuniform grid. This technique has been used to geometrically transform a uniformly spaced constellation into nonuniformly spaced constellation points to cluster points near the center close to each other \cite{BettsTIT1994,KaletTCOM1994}. However, this method only  derives the optimal warping function to reduce the symbol error probability  without considering bit labeling.
Since the labeling plays an important role for designing geometrically-shaped modulation formats, existing symbol-wise based signal shaping methods are incompatible with bit-wise based coded modulation system.

An alternative to 4D SP-based modulation format is to optimize the coordinates of the 4D symbols as non-regularly spaced   signal sets in two polarizations \emph{jointly}. Previous works in this area have mainly investigated the modulation formats design in terms of minimum Euclidean distance \cite{RenaudierECOC2012}. This design criterion has been shown to be suboptimal for the medium signal-to-noise ratio (SNR) regime \cite{BinChenJLT2019}. 
Sensitivity gains were achieved in \cite{Kojima2017JLT,BinChenJLT2019,BinChenPTL2019} by optimizing the points in 4D space and also the corresponding binary labeling, by using the information theoretical performance metric called \gls{GMI} \cite[Sec.~4.3]{Szczecinski2015BICM},  \cite{AlvaradoJLT2018,Alvarado2015_JLT}. \gls{GMI}  can  be  directly  connected  to  modern  binary soft-decision forward error correction (SD-FEC) based on \gls{BICM} \cite{Szczecinski2015BICM,Alvarado2015_JLT}. 
 {Due to its simplicity and flexibility, BICM is usually considered to be a pragmatic approach to \gls{CM} \cite{SmithJLT2012} and hence, the use of \gls{GMI} is  preferred for  optical  fiber communication  systems  design \cite{AlvaradoJLT2018}.}
In  recent experimental demonstrations, the potential of 4D and 8D modulation formats was highlighted. 4D and 8D formats were shown to outperform conventional formats at the same \gls{SE} of 6~bit/4D-sym {\cite{SjoerdOECC2019,LigaPTL2020}} and 5.5~bit/4D-sym, resp. Even though these 4D and 8D formats give interesting performance advantages, larger \gls{GMI} gains (available for larger constellation sizes and higher dimensionality) are  difficult to obtain due to the challenging multi-parameter constellation and labelling optimization. Previous works only solve the 4D or 8D \gls{GMI} optimization problem for up to \gls{SE} of 6 bits/4D-sym \cite{ZhangOFC2017,BinChenJLT2019}.

In this paper, we propose a novel four-dimensional  orthant-symmetric 128-ary modulation format (4D-OS128),  {which has a \gls{SE} of 7~bit/4D-sym and is obtained via GS by jointly optimizing constellation coordinates and labeling in 4D to maximize \gls{GMI}.} For the design we use the orthant symmetry idea  {which can significantly reduce the dimensionality of searching space within the first orthant, and}  overcome the challenging multi-parameter optimization, which can be seen as a trade-off between  {optimization speed and GMI performance}. We found that the obtained 4D format with the orthant symmetry constraint has negligible performance loss with respect to the one without  {orthant symmetry constraint}. The obtained 4D format is compared in terms of linear performance to 128-SP-QAM and a 7 bit modulation in 4D-2A8PSK family (7b4D-2A8PSK) \cite{Kojima2017JLT}, all of them having the same  \gls{SE}  (7~bit/4D-sym).
 {In addition, the proposed 4D-OS128 is also compared to  $\boldsymbol{D}_4$ lattice-based constellation and probabilistically-shaped 16QAM with finite blocklength  via simulation \cite{BinECOC2020}}.
The transmission performance is investigated by  both numerical simulations and experiments. We experimentally  demonstrate two amplifier configurations, EDFA-only and hybrid amplification, with  a data rate of 233~Gb/s per channel. We target a \gls{GMI} lower than the SD-FEC threshold of 5.95~bit/4D-sym \cite{Kojima2017JLT}\footnote{Note that other  thresholds could be used, e.g., the thresholds in \cite[Table~III]{AlvaradoJLT2015}. The use of different thresholds will however not change the general conclusions of reach increase in this paper.}, which corresponds to a FEC rate of 0.8 (25\% FEC OH). For the baseline constellations, distances around 5000~km (EDFA-only) and 8000~km (hybrid amplification) are therefore targeted. Compared at the same bit rate, the proposed {4D-OS128} format achieves a 15\% longer transmission reach than 128SP-16QAM and 7b4D-2A8PSK for 11 \gls{WDM} channels transmission.

This paper is organized as follows.  In Sec.~\ref{Opt}, the design methodology and the proposed modulation format are introduced. In Sec.~\ref{sec:simulation}, numerical results are shown for  both AWGN and nonlinear optical fiber simulations. The experimental setup of the \gls{WDM} optical fiber system and the experimental results are described in Sec.~\ref{sec:experiments}. Conclusions are drawn in Sec.~\ref{con}.

\section{4D Modulation Format and Optimization}\label{Opt}
\subsection{GS Optimization: General Aspects}

The optical channel suffers from the interactions between  \gls{ASE} noise, dispersion and Kerr nonlinearities, which are usually classified
as  inter-channel and intra-channel effects.
This channel can be  modeled by a conditional PDF $p_{\un{\bY}|\un\bX}$, where $\un\bX$ and $\un\bY$ are the transmitted and received sequences, respectively. The transmitted symbols $\bX$  in  $\un\bX$ are assumed to be MD symbols with $N$ real dimensions (or equivalently, with $N/2$ complex dimensions) drawn uniformly from a discrete constellation $\mathcal{X}$ with cardinality $M = 2^m = |\mathcal{X}|$. The most popular case in fiber optical communications is $N = 4$, which corresponds to coherent optical communications using two polarizations of the light (4 real dimensions). This naturally results in four-dimensional (4D) modulation formats. 

In general, the channel law $p_{\un\bY|\un\bX}$  shows memory across multiple symbols, which is introduced by the fiber optical channel even after dispersion compensation.\footnote{One empirical model that properly takes this effect into account for channel capacity calculations is the so-called finite-memory GN model \cite{Agrell2014_JLT}. Another example is the time-domain perturbation models in \cite{DarJLT2015,DarJLT2016,GhazisaeidiJLT2017}, where the received symbol depends on previous and future transmitted symbols (potentially across other polarizations and channels).
 {The memory can also be considered in the demapper over multiple consecutive time slots \cite{ErikssonJLT2017}.}} 
From now on, however, we consider a channel law $p_{\bY|\bX}$ where the output symbols $\bY$ have also $N/2$ complex dimensions. Due to this assumption, we are actually only approximating the true optical channel. As we will explain below, this approximation is well-matched to the fact that typical optical receivers ignore potential memory across 4D symbols.

Throughout this paper, we consider \gls{BICM}, which is one of the most popular \gls{CM} schemes. The transmitted symbols $\bX$ are jointly modulated in 4D space  by a set of  constellation coordinates and the corresponding labeling strategy. The $i$th constellation point is denoted by $\bs_i=\left[s_{i1},s_{i2},s_{i3},s_{i4}\right] \in\mathcal{R}^4$ with $i=1,2,\dots,M$ in four real dimensions. We use the $M\times 4$ matrix  $\mathbb{S}=[\bs_1;\bs_2;\ldots;\bs_M]$  to denote the 4D constellation.\footnote{Notation convention: Random (row) vectors are denoted by $\bX$ and its corresponding realization are denoted by $\bx$. Matrices are denoted by $\mathbb{X}$. A semicolon is used to denote vertical concatenation of vectors, e.g., $\mathbb{X}=[\bx_1;\bx_2]=[\bx_1^T,\bx_2^T]^T$, where $[\cdot]^T$ denotes transpose.} The $i$th constellation point $\bs_i$ is labeled by the length-$m$ binary bit sequence $\bb_{i}=[b_{i1},\ldots,b_{im}]\in\{0,1\}^m$. The binary labeling matrix is denoted by a $M\times m$ matrix $\mathbb{B}=[\bb_1;\bb_2;\ldots;\bb_M]$,  which contains all unique length-$m$ binary sequences. The 4D constellation and its binary labeling are fully determined by the pair of matrices $\{\mathbb{S},\mathbb{B}\}$. 

In this paper, we are interested in  {minimizing the SNR requirements for a target \gls{GMI}, but it can be also  equivalent to maximizing the \gls{GMI} for a given SNR.} The transmitted bits are assumed to be independent and uniformly distributed, which implies uniform symbols $\bX$. The receiver assumes a memoryless channel and also uses a bit-metric decoder (i.e., a standard \gls{BICM} receiver). In this case, the receiver uses a decoding metric $\mathbbm{q}(\by,\bc)$ proportional to the product of the bit-wise metrics, i.e., the decoding metric is
\begin{align}
\vspace{-0.7em}
\mathbbm{q}(\by,\bc) & \propto \prod_{k=1}^{m} p_{\bY|C_k}(\by|c_k)\\
            & \propto \prod_{k=1}^{m}\sum_{b\in\{0,1\}} \sum_{j\in\mcIkb}p_{\bY|\bX}(\by|\bx_{j}),
\vspace{-0.7em}
\end{align}
where $\bC=[C_1,C_2,\ldots,C_m]$ is the random vector representing the transmitted bits, and
$\mcIkb\subset\set{1,2,\ld,M}$ with $|\mcIkb|=M/2$ is the set of indices of constellation points whose binary label is $b$ at bit position $k$.

Under the assumptions above, the \gls{GMI} can be expressed as
\begin{align}
\vspace{-0.7em}
\label{gmi.def0}
G(\mathbb{S},\mathbb{B},p_{\bY|\bX})& = \sum_{k=1}^{m} I(C_k;\bY),
\vspace{-0.7em}
\end{align}
where $I(C_k;\bY)$ is the \gls{MI} between the bits and the symbols, and where the notation $G(\mathbb{S},\mathbb{B},p_{\bY|\bX})$ emphasizes the dependency of the GMI on the constellation, binary labeling, and channel law. Furthermore, for any $N$-dimensional channel law, \eqref{gmi.def0} can be expressed as \cite[eqs.~(17)--(18)]{AlvaradoJLT2018}
\begin{align}
    \nonumber
G(\mathbb{S},\mathbb{B},p_{\bY|\bX})=&m+\frac{1}{M}\sum_{k=1}^{m}\sum_{b\in\set{0,1}}\sum_{i\in\mcIkb}\\
\label{gmi.def}
&\hspace{-9ex}\int_{\mathcal{R}^{N}}p_{\bY|\bX}(\by|\bx_{i}) \log_{2}\frac{\sum_{j\in\mcIkb}p_{\bY|\bX}(\by|\bx_{j})}{\sum_{j'=1}^{M}p_{\bY|\bX}(\by|\bx_{j'})} \, \tnr{d}\by.
\end{align}

As shown in \eqref{gmi.def}, a \gls{GMI} optimization requires a jointly optimization of the 4D coordinates and its binary labeling. A \gls{GMI}-based optimization finds a constellation $\mathbb{S}^*$ and labeling $\mathbb{B}^*$ for a given channel  conditional PDF $p_{\bY|\bX}$
and  energy constraints,  i.e.,
\begin{align}\label{eq:OP_GMI}
\{\mathbb{S}^*,\mathbb{B}^*\} & = \argmax_{\mathbb{S},\mathbb{B}: E[\|\bX\|^2]\leq \sigma_x^2} G(\mathbb{S},\mathbb{B},p_{\bY|\bX}),
\end{align}
where $\sigma_x^2$ represents the transmitted power,  $\mathbb{S}^*$ and $\mathbb{B}^*$ indicate the optimal constellation and labeling, resp. 

Note that for any given channel $p_{\bY|\bX}$, the optimization problem in \eqref{eq:OP_GMI} is a single objective function $G$ with multiple parameter  and constraints. From previous works \cite{ZhangOFC2017,BinChenJLT2019}, it is known that the constellation optimization and \gls{GMI} calculation for large constellations and/or for constellations with high dimensionality is computationally demanding. Therefore, an unconstrained optimization is very challenging. Unconstrained formats also impose strict requirements for the generation of the signals (i.e., high-resolution \gls{DAC}) as well as complex MD detectors. 

\subsection{Orthant-Symmetric (OS) Geometric Shaping Optimization}

To solve the multi-parameter optimization challenges described above and to reduce the transceiver requirements, we propose to impose an ``orthant symmetry" constraint to the $N$-dimensional modulation format to be designed. Our proposed approach makes the MD format to be generated from a 
first-orthant labeled constellation (see \emph{Definition~\ref{def:firstOS}} below). These concepts are defined in what follows, and are based on $N$-dimensional orthants, defined as the intersection of $N$ mutually orthogonal half-spaces passing through the origin. By  selecting the signs of the half-spaces, the $2^N$ orthants available in an $N$-dimensional space can be obtained.

Let $\mathbb{L}_q=[\boldsymbol{l}_1;\boldsymbol{l}_2;\ldots;\boldsymbol{l}_{2^q}]$ with $\boldsymbol{l}_j\in\{0,1\}^q$ and $j=1,2,\ldots,2^q$ denote a $2^q\times q$ \emph{labeling matrix of order $q$}, which contains  all  unique  length-$q$ binary  vectors. Let
$\mathbb{H}_k$ be a ${N\times N}$  {mirror matrix} defined as
\begin{align}\label{eq:mirrormatrix}
\mathbb{H}_k=\left[
\begin{matrix}
(-1)^{l_{k1}} & 0 &\ld 
& 0 \\
0 & (-1)^{l_{k2}} &\ld 
& 0 \\
\vdots	&\vdots	&\ddots 
& \vdots \\
0 & 0 &\ld
& 0 \\
0 & 0 &\ld
& (-1)^{l_{kN}} 
\end{matrix}
\right], 
\end{align}
with $k=1,2,\ldots,2^N$ and $\boldsymbol{l}_k=[l_{k1},l_{k2},\ldots,l_{kN}]$ are the rows of the labeling matrix of order $N$ $\mathbb{L}_N$.

\begin{example}[{Mirror matrices}]\label{Example.1}
For the 1D case ($N=1$), there are only two orthants (an orthant is a ray), the  {mirror matrices} are $\mathbb{H}_1=1$ and $\mathbb{H}_2=-1$.
For the 2D case ($N=2$), there are four orthants  (an orthant is a quadrant), the  {mirror matrices} are
\begin{align}
\mathbb{H}_1&=\left[
\begin{array}{@{~}c@{~}c@{~}}
+1 & 0   \\
0 & +1  
\end{array}\right], 
\mathbb{H}_2=\left[
\begin{array}{@{~}c@{~}c@{~}}
-1 & 0   \\
0 & +1  
\end{array}\right]\\
\mathbb{H}_3&=\left[
\begin{array}{@{~}c@{~}c@{~}}
+1 & 0   \\
0 & -1 
\end{array}\right],
\mathbb{H}_4=\left[
\begin{array}{@{~}c@{~}c@{~}}
-1 & 0   \\
0 & -1 
\end{array}\right].
\end{align}
For the $N=3$ case, there are $2^3=8$ orthants (an orthant is a octant), the eight  {mirror matrices} which can be obtained by \eqref{eq:mirrormatrix}, with $\boldsymbol{l}_j\in\{0,1\}^3$, and $j=1,2,\ldots,8$.
\end{example}

We now define a first-orthant and orthant-symmetric (OS) labeled constellations.
\begin{definition}[First-orthant labeled constellation]\label{def:firstOS}
The pair of matrices $\{\mathbb{T},\mathbb{L}_{m-N}\}$ is said to be a first-orthant labeled constellation if $\mathbb{T}=[\boldsymbol{t}_1;\boldsymbol{t}_2;\ldots;\boldsymbol{t}_{2^{m-N}}]$ is a constellation matrix such that $\boldsymbol{t}_j\in\mathcal{R}_+^{N},\forall j \in \{1,2,\ldots,2^{m-N}\}$ (all entries are nonnegative), and $\mathbb{L}_{m-N}$ is a labeling matrix of order $m-N$.
\end{definition}

\begin{definition}[Orthant-symmetric labeled constellation]\label{def:OS}
The pair of matrices $\{\mathbb{S},\mathbb{B}\}$ is said to be an OS labeled constellation if $\mathbb{S}=[\mathbb{S}_1;\mathbb{S}_2;\ldots;\mathbb{S}_{2^N}]$ is a $2^m\times N$ constellation matrix and $\mathbb{B}=[ \mathbb{B}_1;\mathbb{B}_2;\ldots; \mathbb{B}_{2^N}]$ is a $2^m\times m$ binary labeling matrix, where the constellation matrix $\mathbb{S}$ is constructed via $\mathbb{S}_k=\mathbb{T}\mathbb{H}_k$ with $k=1,2,\ldots,2^N$ and $\mathbb{H}_k$ is given by \eqref{eq:mirrormatrix}, and the labeling matrix $\mathbb{B}$ is such that $\mathbb{B}_k=[\mathbb{O}_k,\mathbb{L}_{m-N}]$, where $\mathbb{O}_k=[\boldsymbol{l}_k;\boldsymbol{l}_k;\ldots;\boldsymbol{l}_k]$ and $\boldsymbol{l}_k$ are the rows of $\mathbb{L}_{N}$, with  $k=1,2,\ldots,2^N$, and $\mathbb{L}_{m-N}$ is a labeling matrix of order $m-N$.
\end{definition}

For an $N$-dimensional OS $2^m$-ary constellation in \emph{Definition~\ref{def:OS}}, each orthant contains $2^{m-N}$ constellation points.  The first $2^{m-N}$ constellation points correspond to the points in $\mathbb{T}=[\boldsymbol{t}_1;\boldsymbol{t}_2;\ldots;\boldsymbol{t}_{M/2^N}]$ (see \emph{Definition~\ref{def:firstOS}}), and are found via $\mathbb{S}_1=\mathbb{T}\mathbb{H}_1=\mathbb{T}$ (see in \emph{Definition~\ref{def:OS}}). The constellation points in $\mathbb{S}_k $ with $k> 1$ are generated by   {``mirroring''} the first-orthant points in $\mathbb{S}_1$ to other orthant via $\mathbb{S}_k=\mathbb{T}\mathbb{H}_k$. The binary labeling for the proposed $N$-dimensional OS formats is such that the constellation points in a given orthant $\mathbb{S}_k$ use the binary labeling $\mathbb{L}_{m-N}$, i.e., $m-N$ bits are used within an orthant. The remaining $N$ bits (which define the matrix $\mathbb{O}_k$ in \emph{Definition~\ref{def:OS}}) are the bits used to select the orthant. To clarify this general definition, we present now two examples. We focus on traditional square $2^m$-ary QAM constellations labeled by the binary-reflected Gary code \cite{AgrellTIT2004} (in one and two polarizations), which are shown to belong to the class of OS labeled constellations.

\begin{figure}[!tb]
    \centering
   \includegraphics[width=0.33\textwidth]{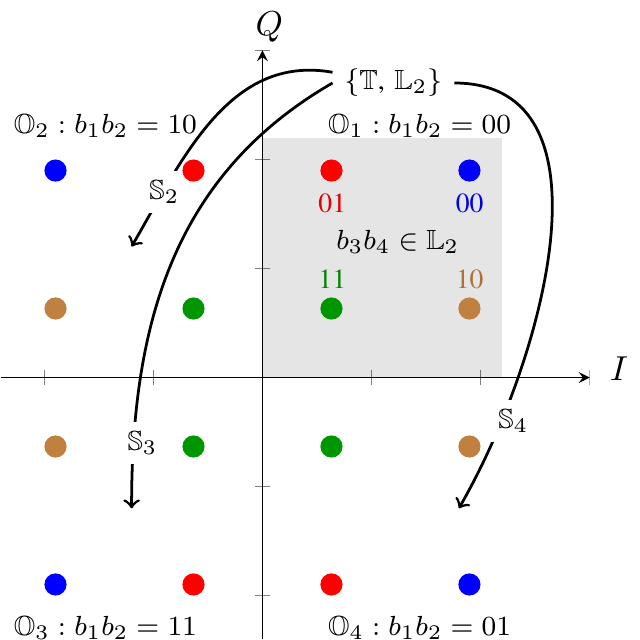}
    \caption{OS labeled constellation example of  16QAM:  $\{\mathbb{T},\mathbb{L}_2\}$ is a first-ortant labeled constellation and $\mathbb{O}_k$  are the bits used to represent the  $k$th orthant. The lines show the mirroring operation to obtain the points $\mathbb{S}_k$ in other  orthant.}
    \label{fig:16QAM}
    \vspace{-1.5em}
\end{figure}

\begin{example}[16QAM] \label{Example.2}
Fig. \ref{fig:16QAM} shows a 16QAM constellation ($N=2$, $m=4$) labeled by the binary-reflect Gray code independently in the first and second dimension. 16QAM is quadrant-symmetric (four quadrants) and consists of the same number of signal points (four points) in each quadrant. The corresponding \textit{first-quadrant (orthant) labeled constellation}  is
\begin{align}\nonumber 
\mathbb{S}_1=\mathbb{T}
=\left[
\begin{array}{@{~}c@{~}c@{~}} \nonumber
3 & 3\\
1 & 3\\
1 & 1 \\
3 & 1
\end{array}\right],
\mathbb{B}_1=[\mathbb{O}_1, \mathbb{L}_{2}]
=\left[
\begin{array}{@{~}c@{}c:c@{}c@{~}} \nonumber
0 & 0 & 0 & 0  \\
0 & 0 & 0 & 1  \\
0 & 0 & 1 & 1  \\
0 & 0 & 1 & 0
\end{array}\right].
\end{align}
The constellation matrix $\mathbb{S}$ and labeling matrix $\mathbb{B}$ of 16QAM  satisfy $\mathbb{S}_k= \mathbb{T}\mathbb{H}_k$ and $\mathbb{B}_k=[\mathbb{O}_k,\mathbb{L}_{m-N}]$ with $k=1,2,3,4$ in Definition \ref{def:OS}, and thus, this format is an OS labeled constellation.
\end{example}

\begin{example}[PM-16QAM]\label{Example.3}
Consider a polarization-multiplexed QAM constellation (PM-16QAM, $N=4$, $m=8$) labeled by the binary-reflect Gray code, independently in each real dimension. This constellation is  obtained as a Cartesian product of two 2D–16QAM, and therefore, PM-16QAM consists of $16$ orthants with $16$ points in each orthant. The \textit{first-orthant labeled constellation} $\{\mathbb{S}_1,\mathbb{B}_1\}$ of PM-16QAM is given by
\begin{align}\nonumber
\mathbb{S}_1  =\mathbb{T} =\left[
\begin{array}{@{~}c@{~}c@{~}c@{~}c@{~}} \nonumber
1 & 1 & 1 & 1   \\
1 & 3 & 1 & 1   \\
3 & 3 & 1 & 1   \\
3 & 1 & 1 & 1   \\
1 & 1 & 1 & 3   \\
1 & 3 & 1 & 3   \\
3 & 3 & 1 & 3   \\
3 & 1 & 1 & 3   \\
1 & 1 & 3 & 3   \\
1 & 3 & 3 & 3   \\
3 & 3 & 3 & 3   \\
3 & 1 & 3 & 3   \\
1 & 1 & 3 & 1   \\
1 & 3 & 3 & 1   \\
3 & 3 & 3 & 1   \\
3 & 1 & 3 & 1   
\end{array}\right],
\mathbb{B}_1  =[\mathbb{O}_1, \mathbb{L}_{4}]
=\left[
\begin{array}{@{~}c@{}c@{}c@{}c:c@{}c@{}c@{}c@{~}} \nonumber
0 & 0 & 0 & 0 & 0 & 0 & 0 & 0  \\
0 & 0 & 0 & 0 & 0 & 1 & 0 & 0  \\
0 & 0 & 0 & 0 & 1 & 1 & 0 & 0  \\
0 & 0 & 0 & 0 & 1 & 0 & 0 & 0  \\
0 & 0 & 0 & 0 & 0 & 0 & 0 & 1  \\
0 & 0 & 0 & 0 & 0 & 1 & 0 & 1  \\
0 & 0 & 0 & 0 & 1 & 1 & 0 & 1  \\
0 & 0 & 0 & 0 & 1 & 0 & 0 & 1  \\
0 & 0 & 0 & 0 & 0 & 0 & 1 & 1  \\
0 & 0 & 0 & 0 & 0 & 1 & 1 & 1  \\
0 & 0 & 0 & 0 & 1 & 1 & 1 & 1  \\
0 & 0 & 0 & 0 & 1 & 0 & 1 & 1  \\
0 & 0 & 0 & 0 & 0 & 0 & 1 & 0  \\
0 & 0 & 0 & 0 & 0 & 1 & 1 & 0  \\
0 & 0 & 0 & 0 & 1 & 1 & 1 & 0  \\
0 & 0 & 0 & 0 & 1 & 0 & 1 & 0  
\end{array}\right].
\end{align} 
It can be shown that the symmetries of PM-16QAM and its binary labeling makes it also to be an OS modulation format.
\end{example}

Now that we have defined OS labeled constellations, we turn our attention back to the GMI optimization in \eqref{eq:OP_GMI}. Because of the OS of the formats under consideration, the new optimization problem is
\begin{align}\label{eq:OP_GMI_simplify}
\{\mathbb{T}^*,\mathbb{L}_{m-N}^*\} & = \argmax_{\mathbb{T},\mathbb{L}_{m-N}: E[\|X\|^2]\leq \sigma_x^2} G(\mathbb{S},\mathbb{B},p_{\bY|\bX}),
\end{align}
whose solution $\{\mathbb{T}^*,\mathbb{L}_{m-N}^*\}$ can then be used to obtain the complete labeled constellation $\{\mathbb{S},\mathbb{B}\}$ using \emph{Definition~\ref{def:OS}}. We emphasize that in \eqref{eq:OP_GMI_simplify}, the GMI is still dependent on the labeled constellation $\{\mathbb{S},\mathbb{B}\}$ (which is fully defined by $2^m$ $N$-dimensional coordinates and a labeling of order $m$), while the optimization is performed only over the $2^{m-N}$ $N$-dimensional coordinates in the matrix $\mathbb{T}$ and the labeling of order $m-N$ (i.e., the matrix $\mathbb{L}_{m-N}$).  {This reduction in the dimensionality of the search space is what will allow us to design a MD format with higher dimensionalities and higher \gls{SE} for maximizing GMI.}

Apart from lowering the optimization complexity, another motivation for using the OS property is that the \gls{GMI}-based constellation optimization can to obtain a Gray-like mapping.
It is well-known that Gray labelings (for which adjacent constellation points differ in only one bit difference) can reduce the loss between \gls{MI} and \gls{GMI} \cite[Fig.~4]{Caire98}, \cite{Agrell10b,AlvaradoTIT2014}, \cite[Sec.~IV]{Alvarado11b}.   {However, since the length of the bit label is shorter than the number of nearest neighbors,
it is not possible to obtain a Gray mapping and also these  constellations also lack an obvious Gray-like mapping.} In order to reduce the non-Gray mapping penalty with respect to  Gray mapping, the orthant-symmetric structure can first guarantee the orthants are Gray-labeled and attempt  to make all the neighbouring symbols in the same orthant with  one bit difference to have a larger distance.

\begin{figure*}[!tb]
\centering
\subfigure[]{ \includegraphics[width=0.31\textwidth]{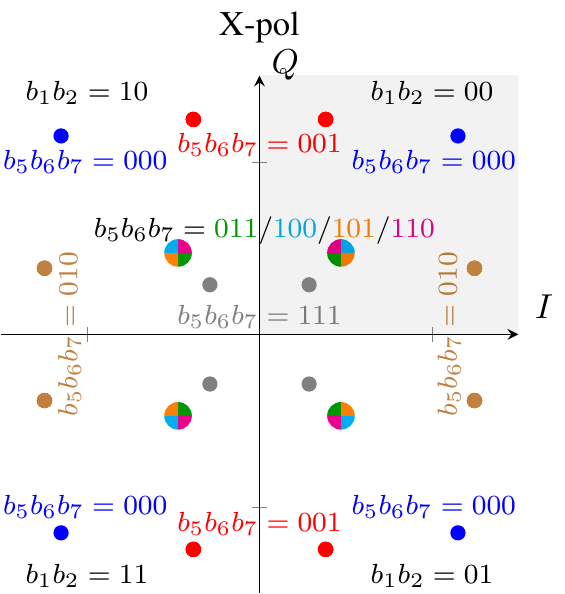}\hspace{0.25em}\includegraphics[width=0.31\textwidth]{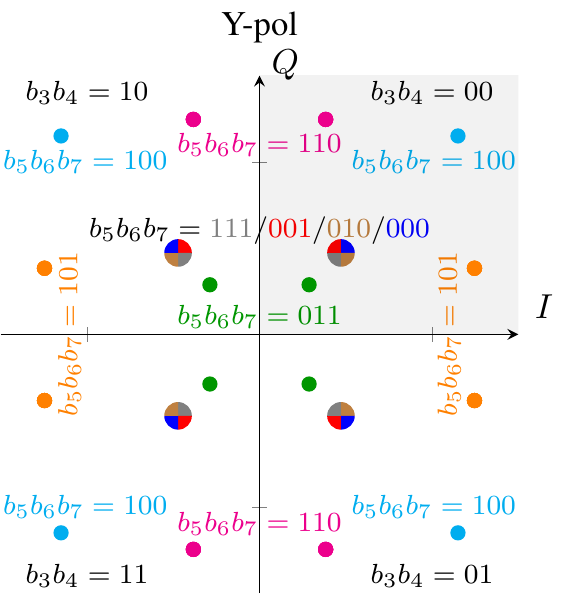}}
\hspace{-0.5em}
\vline
\hspace{-0em}
\subfigure[]{\scalebox{1.03}{\includegraphics[]{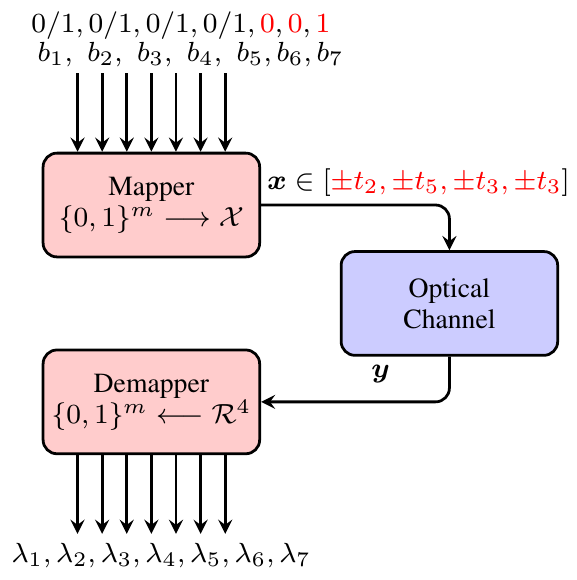}}}
\caption{(a) $2\times$ 2D-projections of the designed 4D-OS128 modulation
and associated bits/polarization
bits mapping. The symbols in the same
color  in two 2D projection indicate a 4D symbol in 4D-OS128 modulation. The shadow area indicate the first orthant in 4D space.
(b) Mapping and demapping example:  sixteen possible 7 bit sequences are mapped to the transmitted 4D symbol $\bx$ (red points in  two 2D projection) with the coordinates $[\textcolor{red}{\pm t_2, \pm t_5, \pm t_3, \pm t_3}]$ and the received 4D symbol $\by$ is  demapped to log-likelihood ratios (LLRs) $\boldsymbol{\lambda}=[\lambda_1,\lambda_2,\lambda_3, \lambda_4,\lambda_5,\lambda_6,\lambda_7]$.} 
\label{fig:4D128}
\vspace{-0.5em}
\end{figure*}

\subsection{Proposed 4D Format: 4D-OS128}

In this   paper, we focus on designing a four dimensional orthant-symmetric 128-ary  (4D-OS128) constellation  ($N=4$ and $M=128$)  with   SEs   of 7~bit/4D-sym ($m=7$), which consists of  $M=2^m$ points $\bs_i,i\in\{1,2,\dots,128\}$ labeled by 7 bits $\bb_i= [b_1,b_2,\dots,b_7]$.
For the 4D-OS128 constellation, each orthant contains $2^{m-N}=8$ constellation points, and therefore the 8 constellation points in first orthant are considered as  the first-orthant  labeled constellation $\mathbb{T}=[\boldsymbol{t}_1;\boldsymbol{t}_2;\ldots;\boldsymbol{t}_8]$ with labeling matrix $\mathbb{G}_{3}$. The $j$th first-orthant  point is denoted by $\boldsymbol{t}_j=\left[t_{j1},t_{j2},t_{j3},t_{j4}\right] \in\mathcal{R}_+^4$ with $j=\{1,2,\dots,8\}$. The first orthant is labeled by four binary bits $[b_{j1},b_{j2},b_{j3},b_{j4}]=\boldsymbol{l}_1$. The remaining 3 bits $[b_{j5}, b_{j6}, b_{j7}]$ determine the point $\boldsymbol{t}_j$ in the corresponding orthant. 

For the 4D-OS128 constellation, only eight 4D coordinates in the matrix $\mathbb{T}$ and the corresponding binary labeling of order 3 $\mathbb{L}_{3}$ need to be optimized in \eqref{eq:OP_GMI_simplify}. 
 {Note that there are many local optima for large constellations and/or for a constellation with high dimensionality,	which has been reported by \cite{ZhangOFC2017}.
To overcome this problem,  we solve the optimization problem \eqref{eq:OP_GMI_simplify}  by applying the approach recently published in \cite{Kadir2019endtoend}, which implement
autoencoders to optimize the constellation  and the binary switch algorithm \cite{Schreckenbach2003}  to find the best ``swap" of two binary labels.}
\footnote{ {Note that initial constellation does affect the time until convergence. In this paper, we use 128SP-16QAM as initial constellation.}}

We optimized the 4D constellation at the target \gls{GMI} of 0.85$m$ bit/4D-sym, where $m$ is the number of bits transmitted per symbol. In other words, the 4D-OS128 modulation was optimized to minimize the SNR requirements for a \gls{GMI} of $0.85\times7=5.95$ bit/4D-sym.
The obtained 4D-OS128 modulation format has 128 nonoverlapping points in 4D space.  For  better  visualization,  these  points  can  be  projected  on  the  two polarizations.
This  projection  results  in  20  distinct  points in  each  2D  space,  as  shown  in Fig.~\ref{fig:4D128} (a).
In  order  to  clearly show  the inter-polarization dependency, we use a similar color coding strategy as in \cite{BinChenJLT2019}: 2D projected symbols in the first and second polarization are valid 4D symbols only if they  share  the  same  color.
The coordinates of the 8 vectors defining the matrix $\mathbb{T}$ are
\begin{align}\label{eq:alphabet}\nonumber
\boldsymbol{t}_j \in \{&[\textcolor{gray}{\pm t_1, \pm t_1, \pm t_3, \pm t_3}], [\textcolor{red}{\pm t_2, \pm t_5, \pm t_3, \pm t_3}],\\\nonumber
&[\textcolor{brown}{\pm t_5, \pm t_2, \pm t_3, \pm t_3}], [\textcolor{blue}{\pm t_4, \pm t_4, \pm t_3, \pm t_3}],\\\nonumber
&[\textcolor{myDarkGreen}{\pm t_3, \pm t_3, \pm t_1, \pm t_1}], [\textcolor{magenta}{\pm t_3, \pm t_3, \pm t_2, \pm t_5}],\\
&[\textcolor{orange}{\pm t_3, \pm t_3, \pm t_5, \pm t_2}], [\textcolor{cyan}{\pm t_3, \pm t_3, \pm t_4, \pm t_4}]\},
\end{align}
where $t_5>t_4>t_3>t_2>t_1>0$. These 8 points are represented in Fig.~\ref{fig:4D128}~(a) with gray, red, brown, blue,  green, magenta, orange  and  cyan  markers in the shadow area, respectively. The  points  with  the  same  color  in  other  orthants  can  be  obtained $\mathbb{S}_k=\mathbb{T}\mathbb{H}_k$, and therefore,  the proposed format is highly symmetric in 16 orthants. 

The  color  coding  scheme  used in Fig.~\ref{fig:4D128}~(a) also  shows  the binary labeling: 4 out of 7 bits [$b_{j1},b_{j2},b_{j3},b_{j4}$] determine the 16 orthant, and the remaining 3 bits [$b_{j5},b_{j6},b_{j7}$] determine the 8 constellation points  in  the  corresponding orthant.  In  other words, [$b_{j5},b_{j6},b_{j7}$]  determines  the  color  of  the  transmitted  points in Fig. \ref{fig:4D128}~(a), while [$b_{j1},b_{j2},b_{j3},b_{j4}$] determine the coordinate of the point in the same color. Fig. \ref{fig:4D128} (b) shows an example of 4D mapping and demapping with  $[b_{j5},b_{j6},b_{j7}]=[0,0,1]$, which indicates that one of the red 4D points in Fig. \ref{fig:4D128} (b) is selected as the transmitted symbol $\bx$.

 {In this paper, the 4D-OS128  modulation  format indicates one of the 4D orthant-symmetric modulations with 128-ary, but optimized for the specific SNR of 9.5~dB.}
The coordinates of the 4D-OS128  modulation  format and  the  corresponding  binary labeling  are  given  in  Appendix (see  Table  \ref{tab:4D_128_XL}).

\subsection{Comparison with Other 4D Modulation Formats}\label{sec:comparebaseline}

\begin{figure}[!tbp]
\centering
\includegraphics[width=0.2\textwidth]{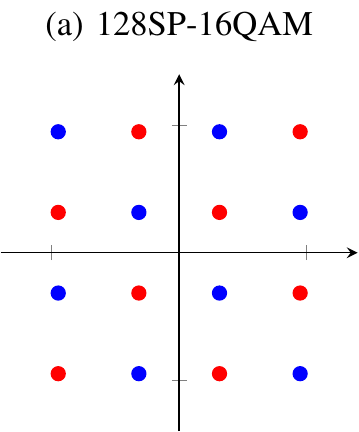}\hspace{1em}\includegraphics[width=0.2\textwidth]{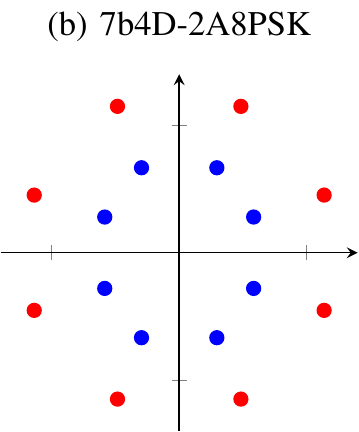}
\vspace{-0.5em}
\caption{2D-projections of  two 4D modulation formats: 128SP-16QAM and 7b4D-2A8PSK. The  colors indicate symbol selecting strategy: the symbols indicated in blue and red are assigned to either of the polarizations.
} 
\label{fig:2A8PSK_128SP16QAM}
\vspace{-0.5em}
\end{figure}

 {Here, we include two other well-known 4D 128-ary formats for comparisons at the same \gls{SE} of 7 bit/4D-sym as GS modulation formats baseline}. The first constellation is 128-ary set-partitioning 16QAM (128SP-16QAM) (see Fig. \ref{fig:2A8PSK_128SP16QAM} (a)), which has been demonstrated for optical communications systems  in \cite{ErikssonOE13,KashiECOC2015,KaihuiOFT2018}. The second one is a 7 bit modulation in the 4D-2A8PSK family (7b4D-2A8PSK) \cite{Kojima2017JLT}  (see Fig. \ref{fig:2A8PSK_128SP16QAM} (b)), which is currently used in  commercial programmable transponders \cite{FujitsuT600}. 
The inner ring/outer ring ratio of   0.59 is chosen for 7b4D-2A8PSK to maximize the \gls{GMI} performance as described in \cite{Kojima2017JLT}.  {Note that 128SP-16QAM and 4D-2A8PSK are set-partitioned from 16QAM and 16-ary two-ring 8PSK, respectively.
128SP-16QAM is obtained by selecting points with larger Euclidean distance. However, for 4D-2A8PSK, constellation points with constant modulus are selected  and the ring ratio of two-ring 8PSK is also optimized for maximizing GMI at the same time.
Therefore, only 7b4D-2A8PSK are designed for an SNR region corresponding to SD-FEC overhead between 20\% and  25\%}

\begin{figure}[!tb]
\centering
\includegraphics[width=0.49\textwidth]{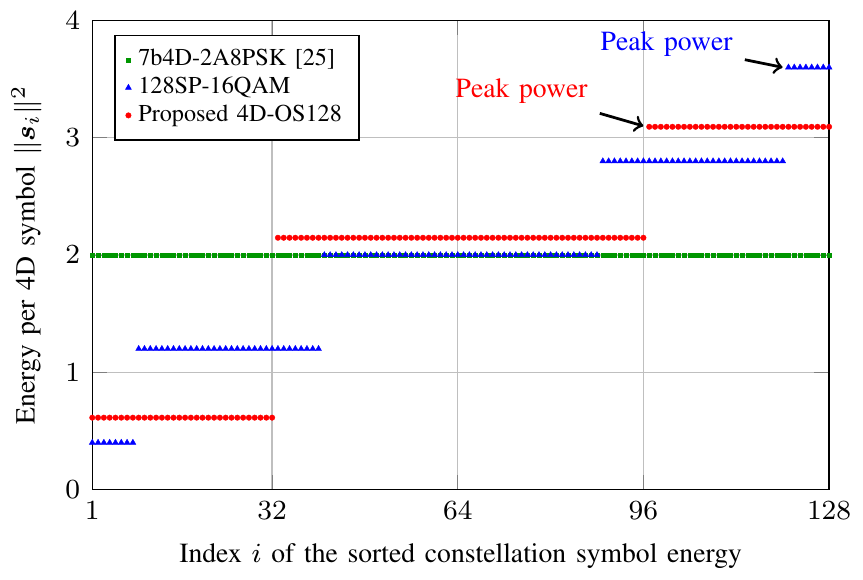}
\vspace{-1.5em}
\caption{Energy per 4D symbol for three \gls{SE} = 7 bits/4D-sym modulations.  The variation of the transmitted symbols'  energy can contribute to the nonlinear interference  noise (NLIN).} 
\label{fig:4D128Energy}
\vspace{-1em}
\end{figure}

To better understand the \gls{GMI} performance and tolerance to fiber nonlinearities, we conduct a  comparison in terms of energy per transmitted symbol, peak-to-average power ratio (PAPR),  variance of  the signal energy with respect to its meanaverage energy $\sigma_{P_s}^2=E\left[\|\bS\|^2-E[\|\bS\|^2]\right]$, 
squared Euclidean distance (SED), and the numbers of pairs of constellation points at minimum squared Euclidean distance (MSED). While PAPR and energy per transmitted symbol can be seen as a rough indication of nonlinearity tolerance  {\cite{GellerJLT2016}}, SED and the number of pairs at MSED can be argued to be performance indicators for the AWGN channel. The analysis and discussion below only gives an intuition on the performance of the proposed format. A precise comparison of these modulation formats will be presented in both the linear and nonlinear channels by numerical simulations (Sec.~\ref{sec:simulation}) and experimental results  (Sec.~\ref{sec:experiments}).

We assume that all the constellations are normalized to $E_s=2$ (i.e., unit energy per polarization). Under this assumption, the comparison of energy for each 4D symbol  {after sorting} for all three modulation formats  are shown Fig. \ref{fig:4D128Energy}. The 7b4D-2A8PSK format has a constant-modulus property, and thus, it can significantly reduce the nonlinear interference  noise (NLIN) \cite{Kojima2017JLT}. For 128SP-16QAM, 5 energy levels are visible, while the proposed 4D-OS128 shows only 3.

{Table \ref{tab:compare} shows four properties of  four modulation formats under consideration. Three of
 them have  SE of 7~bit/4D-sym. PM-16QAM with SE of 8~bit/4D-sym is also compared as baseline.} 
In the first two columns, we use two performance metrics to compare the modulation-dependent nonlinear interference: PAPR and  $\sigma_{P_s}^2$ for a given modulation format. 
Due to the constant-modulus property, both of these two performance metrics are zero for 7b4D-2A8PSK, which is expected to be better than the other two modulation formats in terms of effective SNR. 
{Since PAPR only depends on the few constellation points with largest energy, it cannot reflect the complete nonlinear performance. In contrast,  $\sigma_{P_s}^2$ is the variation of  all the possible transmitted symbols' energy, and thus, smaller $\sigma_{P_s}^2$ should in principle result in higher nonlinear noise tolerance. 
We expect that {the small difference among  4D-128SP-16QAM, 4D-OS128 and PM-16QAM in terms of PAPR and $\sigma_{P_s}^2$} will  contribute  similar modulation-dependent  nonlinear interference, and thus give similar effective SNR.
These predictions will be confirmed in Sec. \ref{sec:simulation} and Sec. \ref{sec:experiments}.}

\begin{table}[!tb]
\caption{Comparison of 7~bit/4D-sym and 8~bit/4D-sym modulation formats.}
\label{tab:compare}
\centering
{\footnotesize
\begin{tabular}{c|c|c|c|c}
\hline

\hline
{ Mod. Formats} & {PAPR [dB]} & {$\sigma_{P_s}^2$} & { $d^2$} & { $n_d$} 
\\
\hline 

\hline
 \multirow{1}{*}{4D-128SP-16QAM}&  \multirow{ 1}{*}{2.55} & 0.645& \multirow{ 1}{*}{0.8} &  \multirow{1}{*}{864}    
 \\
 \hline
\multirow{1}{*}{7b4D-2A8PSK}&  \multirow{ 1}{*}{0}  & 0 & \multirow{ 1}{*}{0.23} &  \multirow{1}{*}{64}  
\\
\hline
\multirow{ 1}{*}{4D-OS128}&   \multirow{ 1}{*}{1.89}  &0.797 & \multirow{ 1}{*}{0.14} &  \multirow{ 1}{*}{16}  
\\
\hline

\multirow{ 1}{*}{{PM-16QAM}}&   \multirow{ 1}{*}{{2.55}}  &{0.643} & \multirow{ 1}{*}{{0.4}} &  \multirow{ 1}{*}{{768}} 
\\

\hline
\end{tabular}
}
\vspace{-1em}
\end{table}

\begin{figure*}[!tb]
\centering
\subfigure[128SP-16QAM]{
\includegraphics[width=0.95\textwidth]{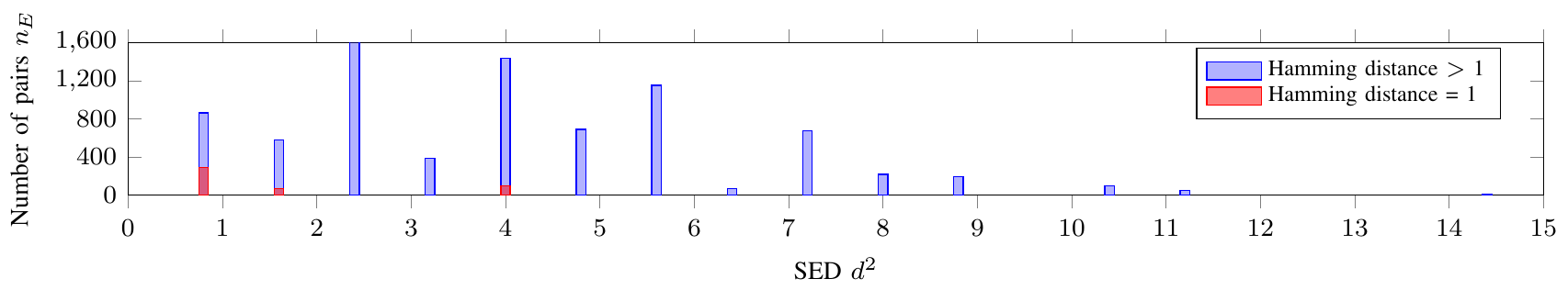}
}
\subfigure[7b4D-2A8PSK]{
\includegraphics[width=0.95\textwidth]{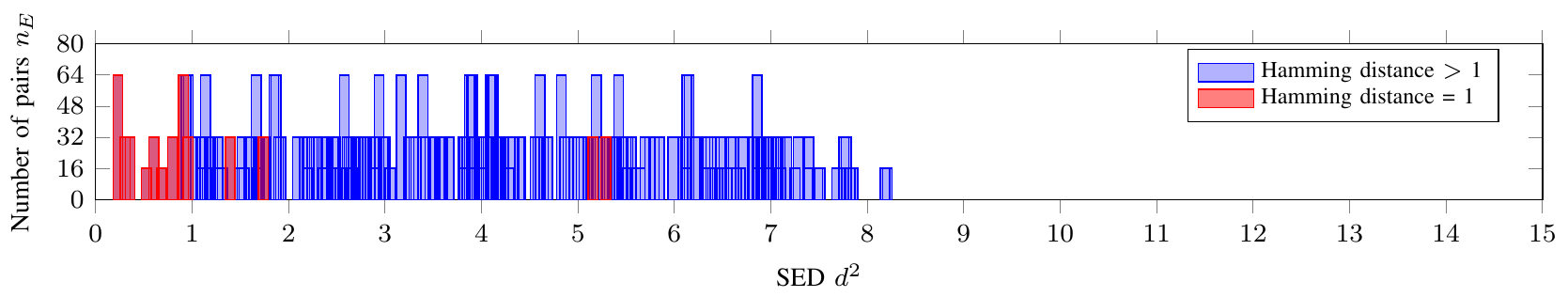}
}
\subfigure[4D-OS128]{
\includegraphics[width=0.95\textwidth]{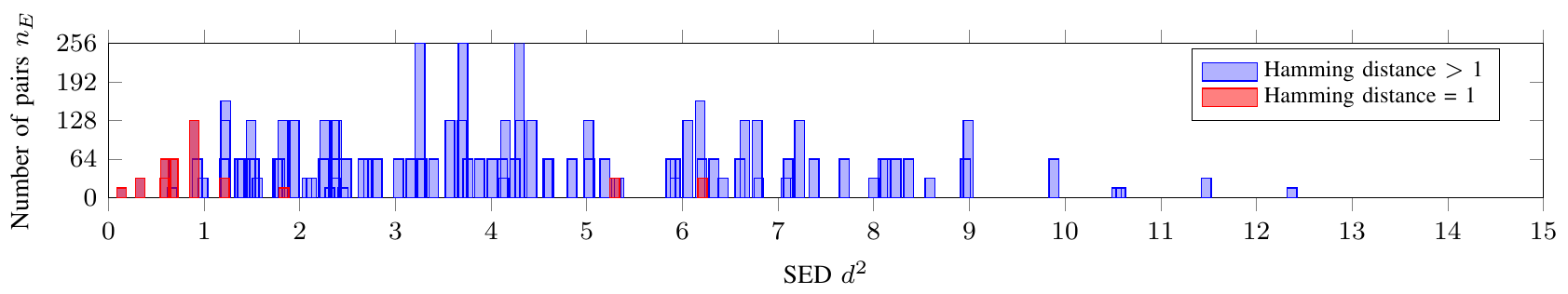}
}
  \caption{Histograms of SEDs of three 4D formats: (a) 4D-128SP-16QAM, (b) 7b4D-2A8PSK, and (c) 4D-OS128. The red bars show the number of pairs with Hamming distance of $1$ at the SED $d^2$. The MSEDs are $0.8$, $0.23$, and $0.14$, for 4D-128SP-16QAM, 7b4D-2A8PSK, and 4D-OS128, resp.}
\label{fig:histSEDs}
\vspace{-1em}
\end{figure*}

In addition to the nonlinear noise tolerance property, we study the structure of the formats in terms of MSED, which we denote by $d^2$. We also look at the number of pairs of constellation points at MSED, which we denote as $n_d$. These two parameters are shown in the last two columns of Table \ref{tab:compare}.
A large $d^2$ and small $n_d$ should in principle result in high MI in the high-SNR regime, as recently proved in \cite{AlvaradoTIT2014}\footnote{The results in \cite{AlvaradoTIT2014} hold for 1D constellations only. However, the authors in \cite{AlvaradoTIT2014} conjectured that the results holds verbatim for any number of dimensions.}. Even though the proposed 4D-OS128 has the smallest $d^2$, it has only 16 pairs at MSED. 4D-128SP-16QAM and 7b4D-2A8PSK have a large number of pairs of constellation points at MSED, which will degrade the  \gls{GMI} performance at medium SNR range.
To better understand this, we also study the SED ``spectrum'' for the three constellations. This is shown as a histograms in Fig.~\ref{fig:histSEDs}. It has been shown in \cite{BinChenJLT2019} that \gls{GMI} does not only depend on $d^2$ and $n_d$, but also the Hamming distance (HDs) of the binary labels of the constellation points at MSED. This figure also shows a classification of the pairs at a given SED: blue bars for pairs at HD larger than one, and red bars for pairs at HD one. From the SED spectra in Fig.~\ref{fig:histSEDs}, we can see that there are less pairs at lower SED and most of the pairs are at Hamming distance one for the proposed 4D-OS128, which in principle results in a better \gls{GMI} in medium SNR range.

\section{Simulation Results}\label{sec:simulation}

\subsection{Linear Performance}\label{sec:linear_simulation}
\begin{figure}[!tb]
\centering
\includegraphics[width=0.485\textwidth]{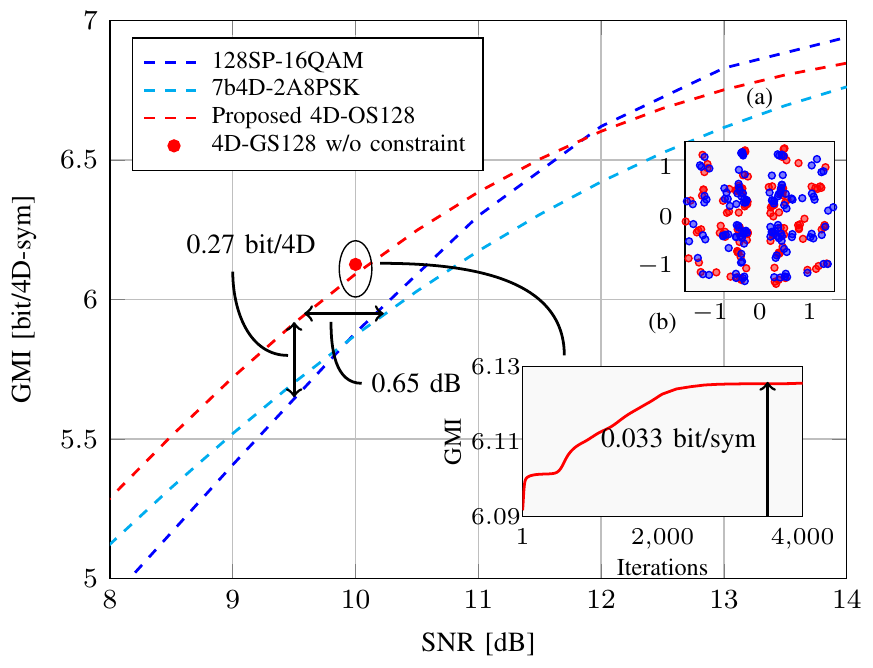}
\vspace{-1.5em}
\caption{GMI vs. SNR for three \gls{SE} = 7 bits/4D-sym modulations. Inset: (a) Optimized constellation of 4D-GS128 w/o orthant-symmetric constraint (b) The optimization procedure after removing orthant-symmetric constraint.} 
\label{fig:4D128GMI}
\vspace{-1em}
\end{figure}

Considering the 128SP-16QAM and 7b4D-2A8PSK as the baselines, Fig.~\ref{fig:4D128GMI}  shows   the  linear  performance  in  terms  of  GMI  for the proposed 4D-OS128 modulation format.
The results  in Fig.~\ref{fig:4D128GMI}  indicates that 4D-OS128 can provide gains of 0.65~dB at GMI of 5.95 bit/4D-sym  {over 128SP-16QAM and 7b4D-2A8PSK}. 
 {The shaping gain comes from the joint optimization of  4D coordinates $\mathbb{S}$ and its binary labeling $\mathbb{B}$  under the constraints of average power and orthant-symmetry.}
Meanwhile, the proposed 4D-OS128 can provide 0.27~bit/4D-sym gain  over  128SP-16QAM at SNR=9.5~dB. 
Eventhough 7b4D-2A8PSK performs better than 128SP-16QAM for SNR below 10~dB, there is at least a gap of 0.5~dB between 7b4D-2A8PSK and 4D-OS128.

In order to verify whether a GMI loss is induced by using the orthant symmetry constraint, we further optimize the 4D 128-ary constellation  by using \eqref{eq:OP_GMI} with 4D-OS128 as an initial constellation   and removing  the  orthant-symmetric  constraint.\footnote{Note that this process is not guaranteed to find the globally optimum constellation.} 
The optimization process can reach steady state within 2000 steps as shown in the inset (b) of Fig.~\ref{fig:4D128GMI}.
The  optimized 4D geometrically-shaped (4D-GS128) constellation without orthant-symmetric constraint is plotted in the inset (a) of Fig.~\ref{fig:4D128GMI}.\footnote{ {Note that 4D-OS128 is also a 4D geometrically-shaped modulation, but with an additional orthant symmetry constraint.}} The red and blue circles represent the symbols transmitted in X and Y polarization respectively.
Despite the 0.033~bit/4D-sym improvement provided by the optimization, the result is a constellation where symbols are very close to each other in the 2D space. This rather complex constellation becomes particularly challenging to generate using a high-speed \gls{DAC} with limited effective number of bits (ENOB).
In this paper, we only investigate the performance of the proposed orthant-symmetric 4D-OS128  modulation  format.

\begin{figure}[!tb]
\centering
\includegraphics[width=0.5\textwidth]{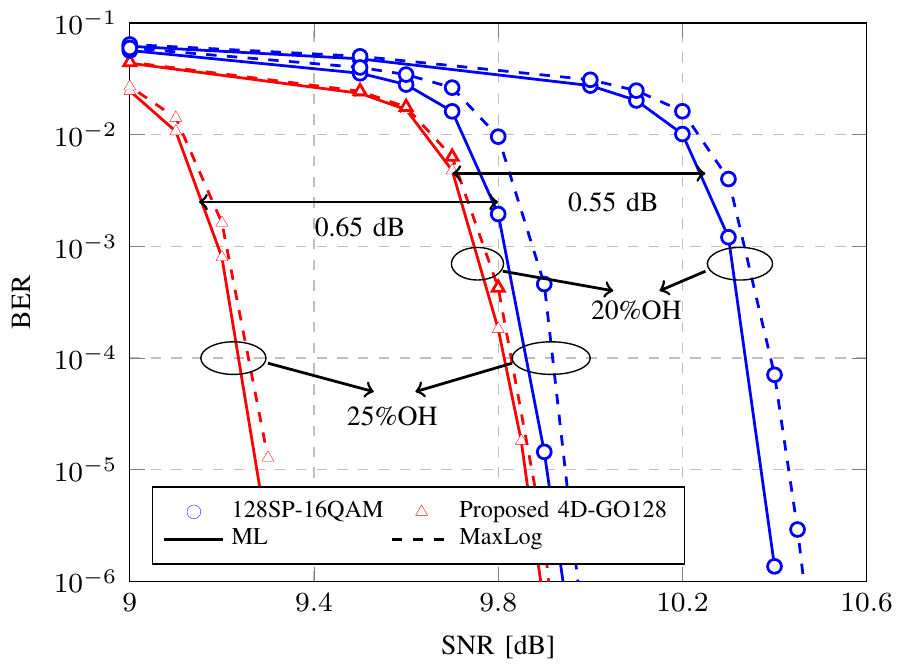}
\vspace{-1.5em}
\caption{Post-FEC BER performance of three \gls{SE} = 7 bits/4D-sym modulations with 20\% and 25\% OH LDPC.} 
\label{fig:4D128BER}
\vspace{-1em}
\end{figure}

For verifying \gls{GMI}  results, we use LDPC codes from the DVB-S2 standard with code rates $R\in\{0.83,0.8\}$ (20\% and 25\% OH) and blocklength $N=64800$. Fig.~\ref{fig:4D128BER} shows post-FEC BER of BER=$4.5\cdot10^{-3}$, between 0.55~dB and 0.65~dB, which is in excellent agreement with the prediction of the \gls{GMI}.
Moreover, we also evaluate the efficiency of the 4D modulations by using low complexity max-log demapper (MaxLog), which significantly  reduces  the computational complexity by avoiding logarithmic and exponential functions as opposed to the optimum  Maximum likelihood (ML) demapper.
Note that using a MaxLog approximation for the proposed 4D-OS128 leads to no  observable  degradation  with  respect  to  the ML demapper. A slightly larger penalty is observed for 128SP-16QAM, in all cases using either 20\% or 25\% OH.

\subsection{Nonlinear Performance: Multi-span WDM transmission}\label{sec:nonlinear_simulation}
We consider a dual-polarization long-haul \gls{WDM} transmission system
 with 11 co-propagating channels generated at a symbol rate
 of  {41.79~GBaud}, a \gls{WDM} spacing of 50 GHz and a \gls{RRC} filter roll-off factor of 0.1. Each \gls{WDM} channel
carries $2^{16}$  4D symbols in two polarizations at the same launch
 power per channel $P_{\text{ch}}$. 
  For the transmission link, a multi-span \gls{SSMF} is used with attenuation $\alpha= 0.21$ dB $\cdot$ km$^{-1}$, dispersion parameter $D = 16.9$ ps$\cdot$nm$^{-1}$ $\cdot$km$^{-1}$, and nonlinear coefficient $\gamma = 1.31$ W$^{-1}$ $\cdot$km$^{-1}$.
 Each span consists of an  {75 km}
 \gls{SSMF} through a split-step Fourier
 solution of the nonlinear Manakov equation with step size
 0.1~km and is followed by an \gls{EDFA}
 with a noise figure of 5 dB.

At the receiver side, channel selection is firstly applied and
then, the signal is downsampled to 2 samples/symbol.  Chromatic dispersion (CD) compensation is performed before applying an RRC matched filter and downsampling
to 1 sample/symbol. An ideal phase rotation compensation is
performed. Then, log-likelihood ratios (LLRs) are calculated and passed to the soft-decision LDPC decoder.
We focus on evaluating the performance of the center \gls{WDM}
channel because it is more affected by the inter-channel crosstalk and non-linear interference.

\begin{figure*}[!tb]
\subfigure[Effective SNR vs. $P_{\text{ch}}$ at  {6750~km}.]{
\includegraphics[width=0.32\textwidth]{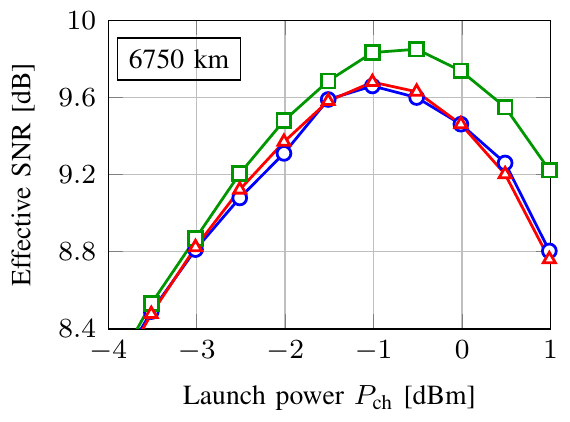}
}
\hspace{-2.5em}
\subfigure[GMI vs. $P_{\text{ch}}$ at  {6750~km}.]{
 \includegraphics[width=0.42\textwidth]{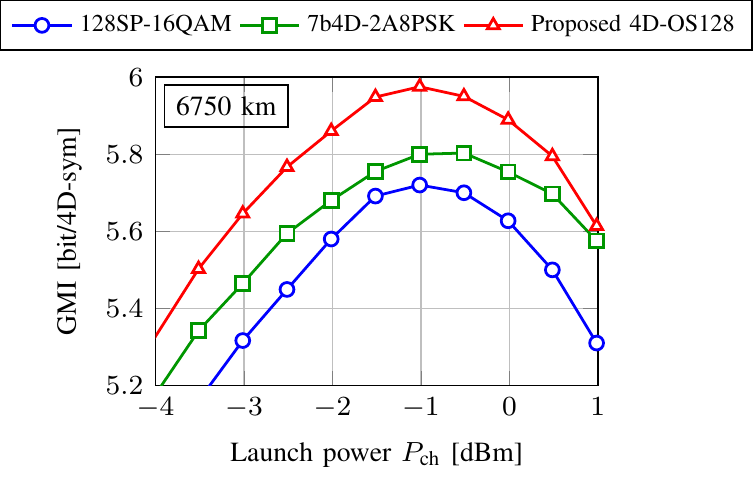}
}
\hspace{-5em}
\subfigure[GMI vs. transmission distance at optimal $P_{\text{ch}}$.]{
  \includegraphics[width=0.33\textwidth]{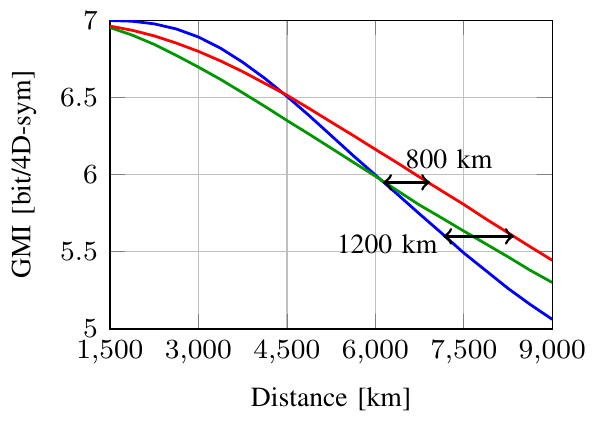}
}
\vspace{-0.5em}
\caption{Simulation results of multi-span optical fiber transmission with 11 WDM channels for three modulation formats with SE of 7~bit/4D-sym: 128SP-16QAM, 7b2D-4D2A8PSK and  4D-OS128.
}
    \label{fig:comparison_7bitmodulation}
\vspace{-1em}
\end{figure*}

Fig. \ref{fig:comparison_7bitmodulation} (a)
shows the effective SNR (after fiber propagation and receiver  DSP) for different 4D modulation formats with SE=7~bits/4D-sym over a  {6750~km }SSMF. 
It is observed that 
the 7b4D-2A8PSK modulation format with constant modulus has less NLI penalty on effective SNR,
 while 128SP-16QAM and the proposed 4D-OS128 with power fluctuation between time slots have an  effective  SNR  penalty  due  to  fiber  nonlinearities. 
Normally, a Gaussian-shaped modulation
format will lead to a larger SNR penalty.
 {However, the proposed 4D-OS128 modulation format experiences a similar effective SNR compared to 128SP-16QAM  at the optimal launch power, which is consistent with the analysis in Sec. \ref{sec:comparebaseline}.}
Therefore, the shaping
gain in the linear regime can be maintained
in nonlinear regime and translates into a reach increase, which will be shown in 
Fig. \ref{fig:comparison_7bitmodulation} (b).

Fig. \ref{fig:comparison_7bitmodulation} (b) shows GMI as a function of the transmitted power for different modulation formats with SE=7 bits/4D-sym over a  {6750~km }SSMF. As we expected, the proposed 4D-OS128 modulation format achieves the highest GMI compared to 128SP-16QAM and 7b4D-2A8PSK. The gains compared to 128SP-16QAM for different launch power are almost constant due to similar effective SNR performance. 
Meanwhile, the gains compared to 7b4D-2A8PSK is reduced as the launch power increases.

Fig. \ref{fig:comparison_7bitmodulation} (c) shows GMI as a function of transmission distance for different modulation formats with SE=7 bits/4D-sym  {using the optimal launch power at each distance}.  {The proposed 4D-OS128 modulation format leads to a 800~km (13\%) and 1200~km (17\%) increase in reach
relative to the 128SP-16QAM modulation format at GMI of 5.95~bit/4D-sym abd 5.6~bit/4D-sym, respectively. In addition, 4D-OS128 perform better than 7b4D-2A8PSK in all the distance.} 
For GMI above 6.5~bit/4D-sym, 128SP-16QAM provides the best performance. This can not only be attributed to the larger  minimum  Euclidean distance for 128SP-16QAM with respect to 4D-OS128 (see Table \ref{tab:compare}). In addition,  4D-OS128 is not designed for higher SNR at shorter transmission distances.

\subsection{4D-OS128 vs. Other Formats and Probabilistic Shaping}

{In the previous section,  the proposed 4D-OS128 modulation  comparable to 128SP-QAM and 7b4D-2A8PSK were discussed for BICM, which enables low complexity.
To close the gap to Shannon limit, an alternative approach would be the use of  multi-level coding (MLC) with multi-stage
decoding (MSD) together with constellations based on  the $\boldsymbol{D}_4$ lattice.
Recently, 4D \gls{CM} based on
the set of Hurwitz integers has been proposed and demonstrated targeting the (IA)  OIF 400ZR implementation agreement   \cite{FreyECOC2019,FreyJLT2020}.
The  cardinality of the resulting Hurwitz constellation is   $M=2\cdot2^{4i}= 32, 512, 8192,\dots$, where $i$ is an integer.
By selecting the set of Hurwitz integers from the $\boldsymbol{D}_4$  lattice, it is not possible to construct a   constellation with $M=128$.
Therefore, we consider here a   lattice-based constellation with \gls{SE} of 7~bit/4D-sym instead, which we denote by $l_{4,128}$.
The constellation is the optimal spherical subset of $\boldsymbol{D}_4$ and   thus by enumerating and testing a finite number of possible centroids inside the fundamental simplex of the lattice, it becomes optimal.
The $l_{4,128}$ constellation was first characterized in \cite[Fig. 1(b)]{KarlssonOFC2012} and  it also
corresponds to the  format  ``l4\_128"in \cite{ErikDatabase}.
}

\begin{figure}[!tb]
\centering
\includegraphics[width=0.49\textwidth]{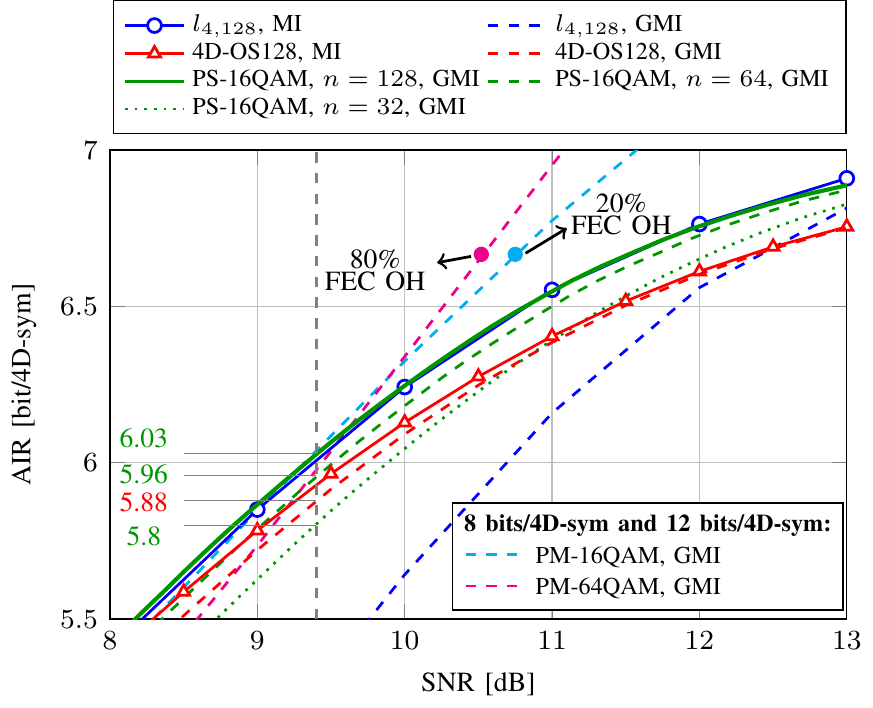}
\vspace{-2em}
\caption{AIR vs. SNR for different modulation formats with SE of 7~bit/4D-sym. {PM-16QAM  with SE of 8~bit/4D-sym and  PM-64QAM with SE of 12~bit/4D-sym are also shown as baselines.} }
\label{fig:AWGN_SNRvsNGMI_comaprision}
\vspace{-1.5em}
\end{figure}

 {
To label the constellation $l_{4,128}$, we use a numerically optimized
labeling obtained using the binary-switching algorithm \cite{Schreckenbach2003}. A labeling was obtained and  optimized for an SNR of  $9.5$~dB.
In Fig. \ref{fig:AWGN_SNRvsNGMI_comaprision}, the \gls{MI} and GMI for the $l_{4,128}$ constellation  are shown.
 {For the proposed 4D-OS128 format, the GMI and the MI are almost identical due to the obtained optimized labeling. 
This is not the case for the constellation $l_{4,128}$, which  lacks a Gray labeling  due to the large amount of nearest neighbors.}
The results in Fig. \ref{fig:AWGN_SNRvsNGMI_comaprision} shows that $l_{4,128}$ gives a high \gls{MI} at all SNRs.
However, achieving MI requires symbol-wise decoders, such as  MLC-MSD, which have often been avoided in optical communications because of the potentially high complexity induced by using separate bit-level codes and the negative impact of decoding delay.\footnote{ {Note that the complexity and delay of MLC-MSD schemes  can be reduced by wise choice of bit-level codes and efficient hardware implementation designs \cite{BarakatainJLT2020}.}}
It is  also known that MLC schemes are generally more sensitive than BICM to a mismatch between the actual channel parameters and those for which the codes are designed \cite{Szczecinski2015BICM}.
Furthermore,  for the $l_{4,128}$ constellation,  a large gap between the \gls{MI} and GMI exists (more than 1~dB for low code rates). 
Therefore, $l_{4,128}$ will not work well with a bit-wise decoder.
Similar results have been previously reported in \cite{AlvaradoJLT2015} for 4D formats with $M=16, 256, 4096$.
}

{
In Fig. \ref{fig:AWGN_SNRvsNGMI_comaprision},   PM-16QAM  with  SE  of  8~bit/4D-sym  and  PM-64QAM  with  SE  of 12~bit/4D-sym are also shown as baselines.
We observe that PM-64QAM  has higher GMI at higher SNR region (SNR$\geq$10~dB) compared to PM-16QAM.
However, the comparison need to be made between  different FEC overheads.
For example,  PM-64QAM with 80\%  FEC OH can outperform PM-16QAM with   20\%  FEC OH at GMI of 6.67~bit/4D-sym, at the price of higher decoding complexity.
The same argument applies to the comparison between PM-16QAM and all other formats with SE of 7~bit/4D-sym.
More details about  higher-order modulations with lower-rate FEC  are discussed in Sec. \ref{sec:16QAM_FEC_com}.}

\begin{figure*}[!tb]
	\subfigure[Effective SNR vs. $P_{\text{ch}}$ at 6750~km.]{
	\includegraphics[width=0.32\textwidth]{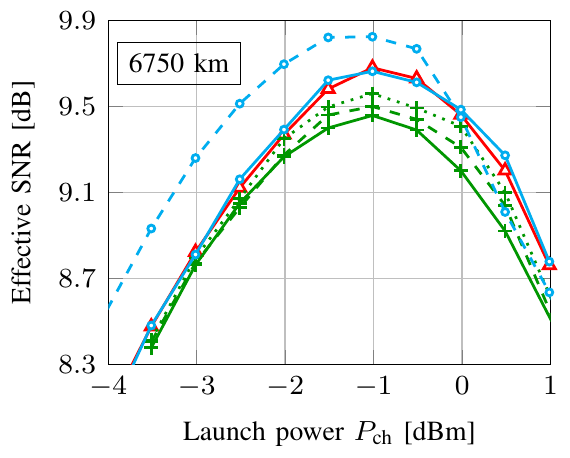}}
	\hspace{-14em}
	\subfigure[GMI vs. $P_{\text{ch}}$ at 6750~km.]{
	\includegraphics[width=0.83\textwidth]{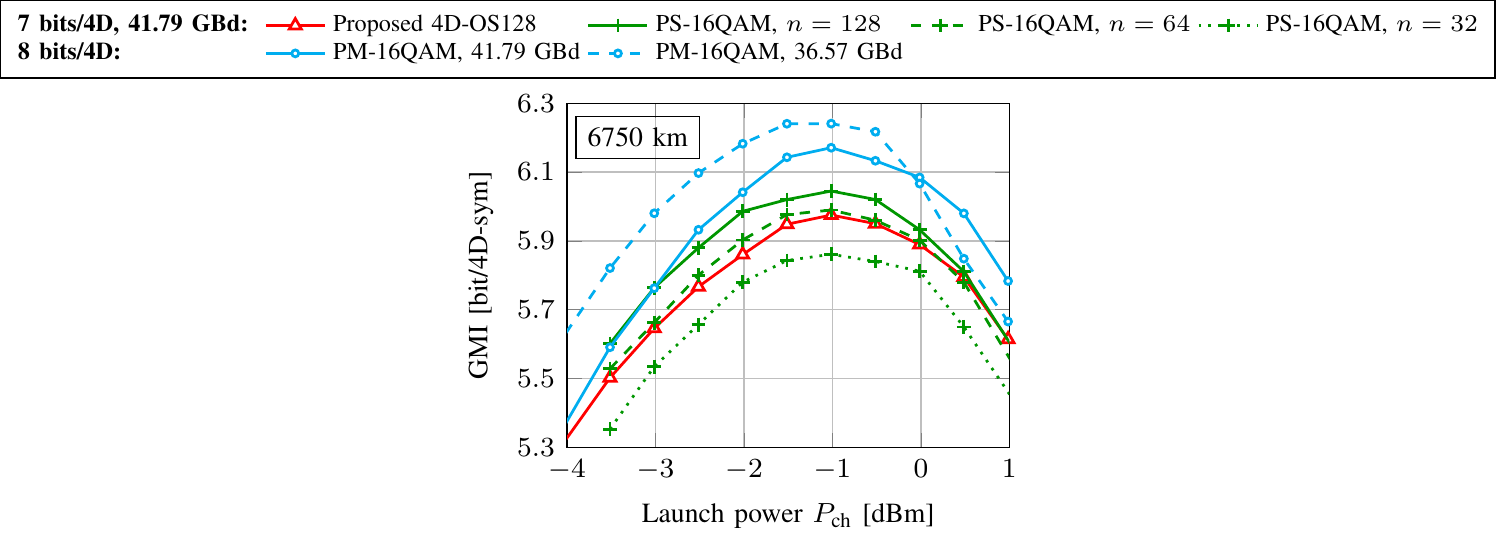}}
	\hspace{-14em}
	\subfigure[Net data rate vs. distance at optimal $P_{\text{ch}}$.]{
	\includegraphics[width=0.335\textwidth]{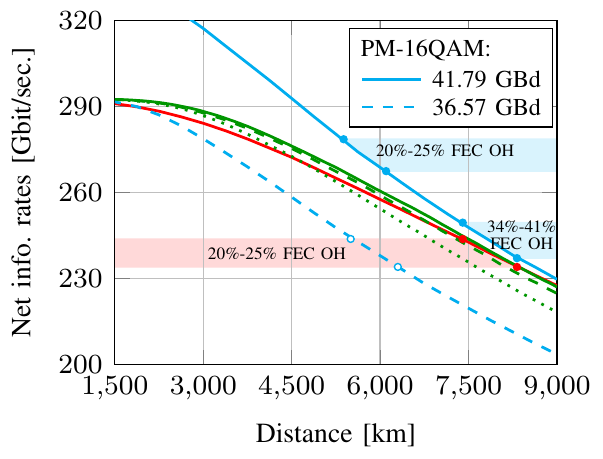}}
	\vspace{-0.5em}
	\caption{Simulation results of multi-span optical fiber transmission with 11 WDM channels for 4D-OS128 and PS-16QAM with three difference blocklengths. The same simulation setup is implemented as Sec. \ref{sec:nonlinear_simulation}.   
		{Results for PM-16QAM with reduced symbol rate (36.57~Gbaud) are also shown, which give the same net data rate at a given coding rate when compared to the other formats. The net data rate results in bits per second are   shown in (c).}}
	\label{fig:comparison_PS}
	\vspace{-1em}
\end{figure*}

{In flexible transponders, different modulation formats are used to target different transmission distances. Often in these transponders, the same FEC is used as a separate subsystem for different formats. 
In order to make a fair comparison against PM-16QAM (8~bits/4D-sym) for the same coding rate, we followed the  methodology  in \cite{ErikssonOE13,KashiECOC2015}, i.e., we reduce the symbol rate by 1/8 (from 41.79 GBd to 36.57~GBd). This symbol rate reduction makes the net data rate of the considered equal for any given coding rate. Results for PM-16QAM for the same symbol rate of 41.79 GBd are also included for completeness.}

{The effective SNR vs. launch  power and GMI vs. launch  power for  a  transmission  distance of  6750~km are  shown in  Fig. \ref{fig:comparison_PS} (a) and  Fig \ref{fig:comparison_PS}  (b), respectively.
We can observe that PM-16QAM with the same symbol rate  of 41.79~GBd  experiences  a  similar  effective SNR compared to the  proposed 4D-OS128, and thus has a higher GMI at optimal launch power due to its better  linear performance.
By contrast, PM-16QAM with 36.57~GBd gives higher effective SNR  due to the larger energy per symbol, and thus also achieves higher GMI at optimal launch power.
}

{Transmission distance vs. net data rate for the optimal launch power are shown in Fig. \ref{fig:comparison_PS} (c). In addition, the required FEC OHs are also highlighted by markers.  
We  observe that PM-16QAM with 41.79~GBd  achieves higher net data rate  than all other  modulation formats in Fig. \ref{fig:comparison_PS} (c). However,  this comes at the cost of higher FEC OH at the same transmission distance.
For example, for the  distances  between 7400~km and 8300~km, PM-16QAM requires an  FEC OH of $34\%-41\%$ and 4D-OS128  only need the FEC OH of $20\%-25\%$.
When the same FEC OH of $20\%-25\%$ is considered, PM-16QAM with 41.79~GBd  could achieve even higher net data rate, but at the cost of shorter transmission distance  (between 5400~km and 6100~km).
}

In addition to multidimensional modulation formats, we also  consider PS-QAM with the same \gls{SE} as baseline, which  probabilistically shapes   PM-16QAM to an entropy of $H(X) = 7$~bit/4D-sym.
It is a well known fact that PS has superior \gls{AIRs} performance for a finite number of constellation points with respect to GS, however, this is based on ideal assumptions for PS, requiring further in-depth analysis;

\begin{itemize}
 \item \textbf{Short blocklength rate loss.} PS based on constant composition distribution matching (CCDM) with long blocklengths is very difficult to implement in high speed	communications because it is based on sequential arithmetic coding. 
  In this paper, we consider the finite-length information rate (defined in 
\cite[eq. (15)]{2018Tobias_PBDM}) 
$\text{AIR}_{n}$ of PS with CCDM blocklength of $n$ as  $\text{AIR}_{n}=\text{GMI}-NR_{\text{loss}}$, where $N$ is the number of real  dimensions and $R_{\text{loss}}$ is the  rate loss of CCDM with a finite blocklength.
The rate loss is defined in \cite[eq. (4)]{2018Tobias_PBDM} as    	\begin{equation}
    	R_{\text{loss}}= H(P_A) -\frac{k}{n}~\text{[bits/amp]},
\end{equation}
where $P_A$ is the targeted probability distribution, $H(P_A)$
is the entropy in bits/amp, $k$ is the number of input bits
for CCDM  and $n$ is the blocklength. 
Note that both PS and GS  coincide, when a block code and low-dimensional constituent constellations is used. 
Thus,  a fair comparison would be to use probabilistic shaping with a blocklength of $n = 4$ as baseline.
But considering the complexity of  demapping for 4D GS modulation, we use $n = 128, 64, 32$ for PS in this paper. 
A more comprehensive complexity comparison is left for future research.
In Fig. \ref{fig:AWGN_SNRvsNGMI_comaprision}, we show the GMIs of PS-16QAM with shaping blockslength $n = 128,64, 32$. 
We observe that PS gives higher GMI with $n=128$, but the resulting rate loss diminishes the efficiency of CCDM as the blocklength decreases.
With  $n = 32$, PS-16QAM has even worse GMI performance than 4D-OS128.
\item \textbf{Lower tolerance to fiber nonlinearities}. 
As  mentioned in \cite{TobiasJLT16}, PS-16QAM experiences higher nonlinearity penalties, which will also lead to a decrease in effective SNR  and GMI.
These losses are shown in Fig. \ref{fig:comparison_PS} (a) and (b), and  are particularly  visible in the high nonlinear regime.
In Fig. \ref{fig:comparison_PS} (a), we also observe that PS with short blocklengths can also slightly increase the nonlinear tolerance, and thus, the effective SNR. The phenomena is  also reported in \cite{Amari2019_IntroducingESSoptics}.
\end{itemize}

By considering the  two aspects discussed above, we show results of  net data rate with the corresponding baud rate as a function of the transmission distance
in Fig. \ref{fig:comparison_PS} (c). We can see that the theoretically superior performance of PS-16QAM vanishes as the blocklength $n$ decreases. 
In addition, for optical links with  stronger nonlinearity, PS will have even higher penalty and lead to a reduced transmission reach and GMI.

When compared to other formats, the proposed 4D-OS128 modulation format has its own advantages in terms of  complexity and performance trade-off.
The 4D-OS128 modulation format can be easily coupled with FEC and only requires straightforward modifications of the mapper and demapper, and thus, it could be an alternative candidate for optical transmission systems.

\subsection{ {Comparison with PM-16QAM with identical FEC decoding complexity}}\label{sec:16QAM_FEC_com}

Normally, modulation formats with the same \gls{SE} are compared with the same FEC code rate to keep the identical FEC decoding complexity.
For comparison of modulation formats with different \gls{SE}, different FEC rates need to be considered.
However, it has been shown in the   literature {\cite{SugiharaOFC2016,Koike-AkinoJLT2016,Koike-AkinoJLT2017}} that higher-order modulations with  lower-rate FEC incur a higher penalty {in terms of complexity} because more decoding iterations per information bit are required to converge.  {When the decoding complexity is constrained, the penalty will be transformed into a rate loss between the information rates of realistic hardware-implementable FEC codes and idealistic BICM limits (GMI)}.

{For belief-propagation based decoding,}
 {the decoding complexity of LDPC codes for each symbol can be approximated to be proportional to the number of belief message updates per information bit as follows \cite[Eq. (2)]{Koike-AkinoJLT2017},\footnote{{Here we only compare the complexity for FEC decoding. However, 4D formats also increase the computational complexity of  demapper with respect to a conventional QAM demapper. Low-complexity 4D
 soft demapper  can be designed to reduce the number of MD Euclidean distances calculation by  utilizing the property of orthant-symmetry and the schemes in \cite{YoshidaECOC2016,BendimeradECOC2018,NakamuraJLT2018}.}}
 
\begin{equation}\label{eq:BP_complexity}
P \propto  \frac{N\bar{d_v}}{R}\cdot \log_2M,
\end{equation}	
where $N$ and $\bar{d_v}$ denote the number of decoding iterations and average
variable-node degree, respectively.  The last term
of $\log_2M$ is  the total number of bits per symbol for different modulation formats.
} {Hence, when comparing different modulation formats at the same bit rate and for a given number of decoding iterations, high order modulation formats result in higher complexity.}

\begin{figure}[!tb]
\includegraphics[width=0.49\textwidth]{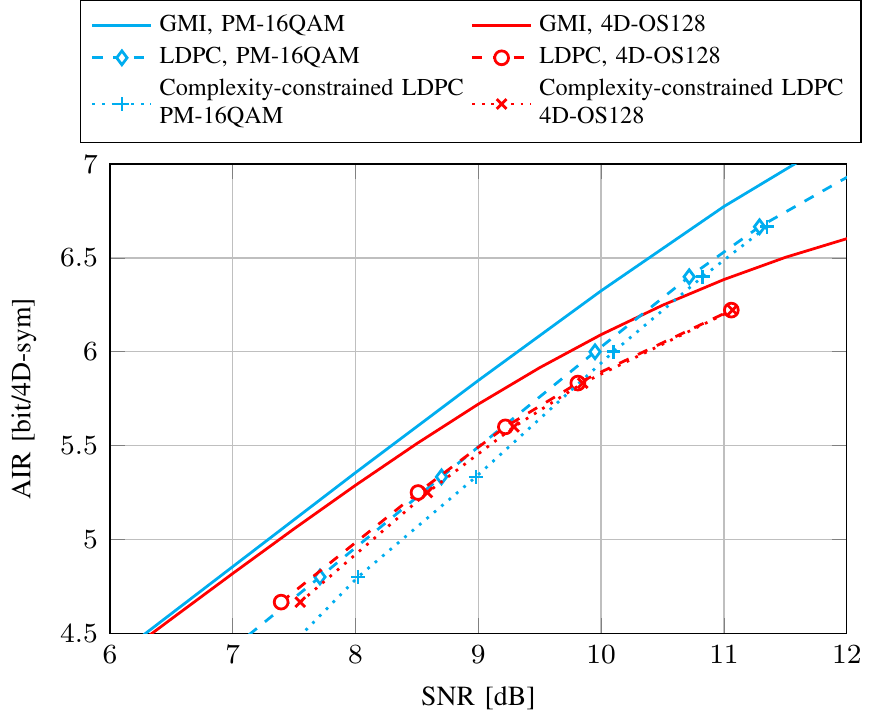}
\vspace{-1.5em}
\caption{AIR vs. SNR for the AWGN channel. The results obtained using LDPC codes are shown with markers, which is the required SNR to achieve a post-FEC bit error probability below $10^{-4}$. {Note that BERs below $10^{-15}$ can be achieved by for example concatenating an outer high-rate Bose–Chaudhuri–Hocquenghem (BCH) code. This code will only add a minimal  additional overhead at an input BER below $10^{-4}$.}}
\label{fig:GMI_SNR_FEC_comparison}
\vspace{-1em}
\end{figure}

{
To keep approximately constant decoding power consumption per symbol as shown in \eqref{eq:BP_complexity}, the LDPC codes with  $R=5/6$, $\bar{d_v}=4$, and $N=50$ iterations for $M=4$ QAM are considered as baseline. Hence,  the number of decoding iterations of complexity-constrained LDPC is adjusted for PM-16QAM and 4D-OS128 with different code rate $R\in\{1/2, 3/5, 2/3, 3/4, 4/5, 5/6, 8/9\}$.
Fig. \ref{fig:GMI_SNR_FEC_comparison} shows the achievable information rates as a function of required SNR of those LDPC codes.  It is observed that achievable rates with LDPC codes still have approximately 1--2~dB loss from GMI, and also that higher-order modulation and lower-rate LDPC codes have higher penalty.
It is interesting to note that while 4D-OS128 cannot outperform PM-16QAM for any SNR in terms of GMI (ideal FEC), it does perform equally well (if not better) for an SNR below 9.5 dB when realistic LDPC codes are used. This suggests that idealistic GMI analysis can be too optimistic to evaluate information rates for the hardware with power consumption limitations.
}

\begin{figure*}[!tb]
\centering
\includegraphics[width=0.93\textwidth]{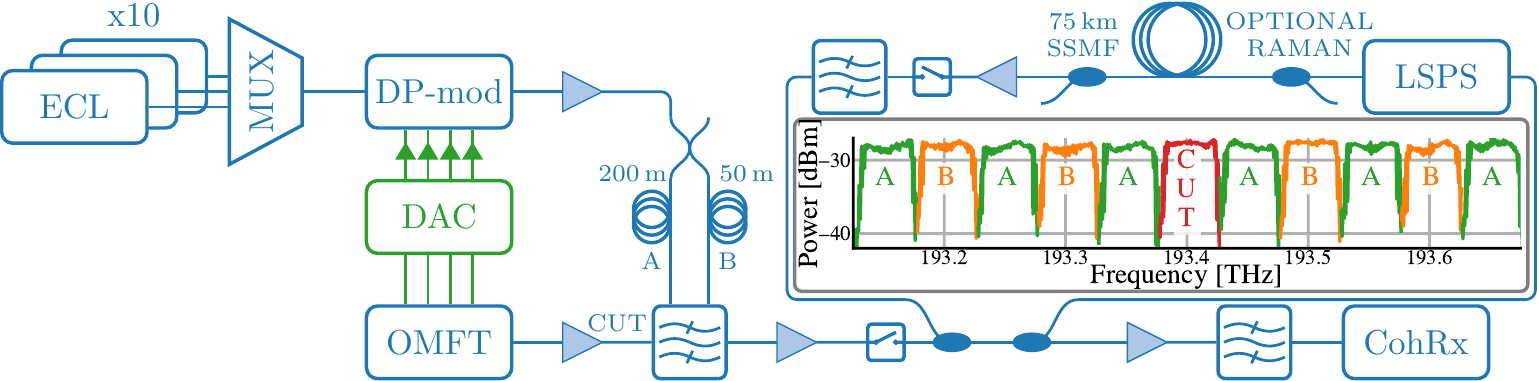}
\caption{Experimental optical recirculating loop setup. Inset: The received spectrum after $N=80$ circulations (6000~km) of EDFA-only amplification. The CUT is depicted in the center position but is tested in all 11 positions in the experiment.}  
 \label{fig:setup}
\end{figure*}

\section{Experimental Setup and Results} \label{sec:experiments}

\subsection{Experimental Transmission Setup}
Fig. \ref{fig:setup} depicts the experimental transmission setup. The transmitted signal is modulated using either 128SP-16QAM, 7b4D-2A8PSK \cite{Kojima2017JLT}, or 4D-OS128 symbols. 
Pseudo-random sequences of 2\textsuperscript{16} symbols are generated offline, pulse shaped using a \gls{RRC} filter with 1\% roll-off at 41.79~GBd, and uploaded to a 100-GSa/s \gls{DAC}. The positive differential \gls{DAC} outputs are connected to the \gls{OMFT} which consists of an \gls{ECL}, a \gls{DPIQ}, an \gls{ABC} and RF-amplifiers. The \gls{CUT}, which can be defined at any of the 11 tested C-band channels, is modulated by the \gls{OMFT} and subsequently amplified. The loading channels are provided by the negative outputs of the \gls{DAC} and modulated onto the tones provided by 10 \glspl{ECL} using a \gls{DPIQ}. These loading channels are amplified, split into even and odd channels, decorrelated by 10,200 (50~m) and 40,800 symbols (200~m), and multiplexed together with the \gls{CUT} on a 50-GHz grid using an \gls{OTF}. Bandwidth limitations due to transmitter electronics are initially compensated using an \gls{OTF} and the residual effects are mitigated digitally as proposed in \cite{LinECOC2018}.

\begin{figure}[!tb]
 \vspace{-0.5em}
\centering
\hspace{-1.5em}
\subfigure[128SP-16QAM]{
\includegraphics[width=0.15\textwidth]{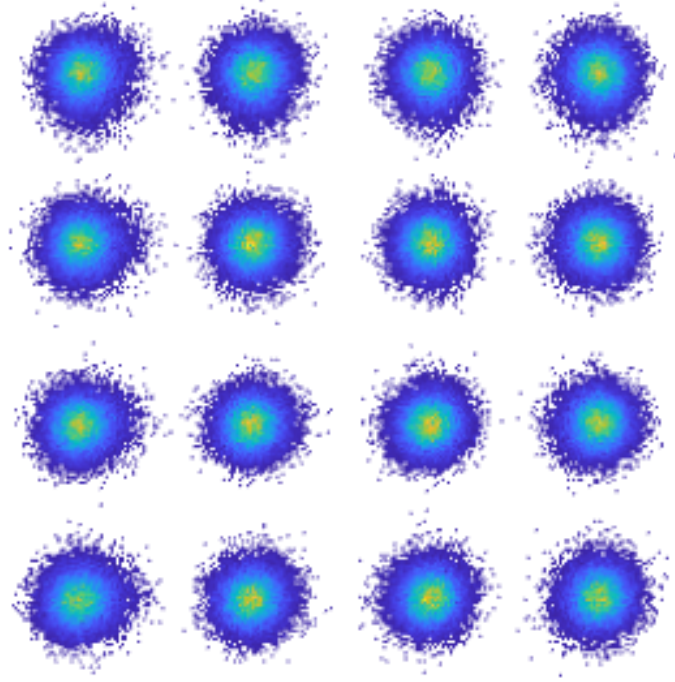}
}
\subfigure[7b4D-2A8PSK]{
\includegraphics[width=0.15\textwidth]{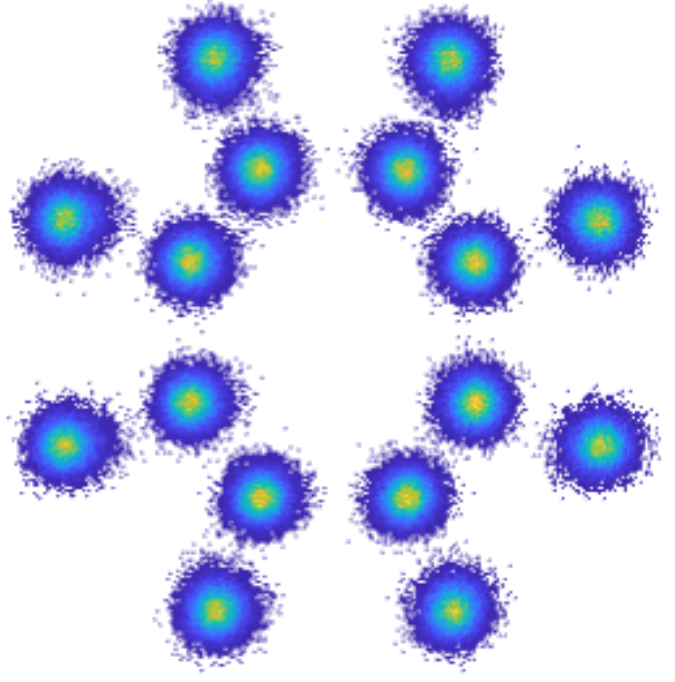}
}
\subfigure[4D-OS128]{
\includegraphics[width=0.15\textwidth]{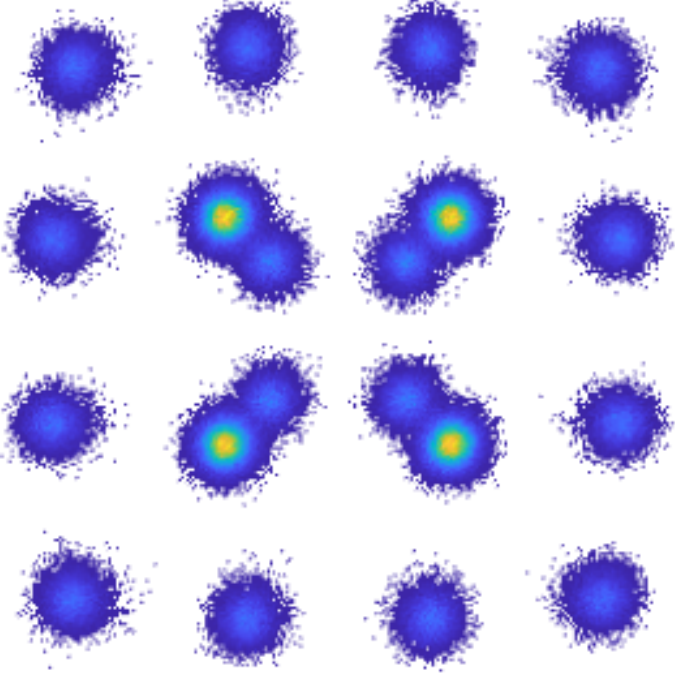}
}
 \vspace{-0.5em}
\caption{2D projection of the constellations after Back-to-Back measurements. The SNR of these recovered constellations is around 20~dB.}
 \label{fig:constellation_B2B}
\end{figure}

\begin{figure}[!tb]
 \vspace{-0.5em}
\centering
\hspace{-1.8em}
\subfigure[128SP-16QAM]{
\includegraphics[width=0.176\textwidth]{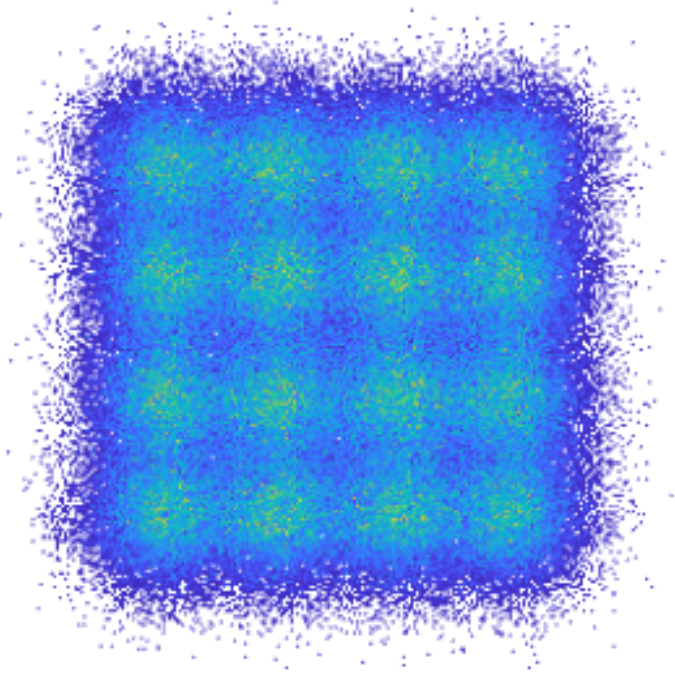}}\hspace{-1.2em}
\subfigure[7b4D-2A8PSK]{
\includegraphics[width=0.176\textwidth]{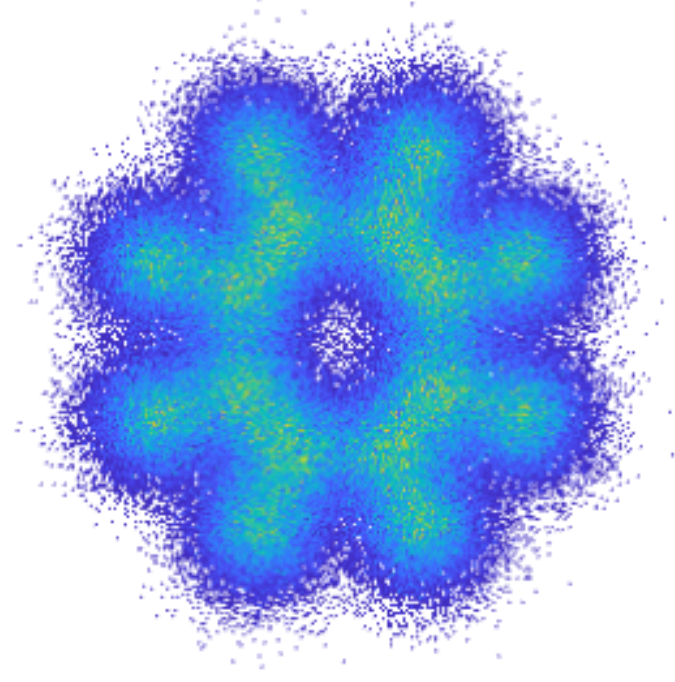}}\hspace{-1.1em}
\subfigure[4D-OS128]{
\includegraphics[width=0.176\textwidth]{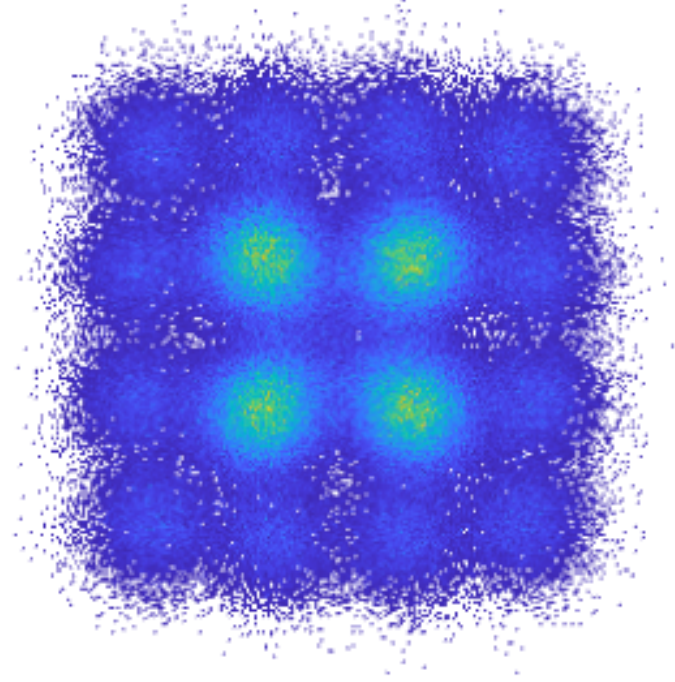}}
 \vspace{-0.5em}
\caption{2D projection of the constellations after  70 spans transmission with  {total} launch power of 9~dBm.}
 \label{fig:constellation_70spans}
 \vspace{-0.5em}
\end{figure}

The 11-channel 50-GHz-spaced \gls{DWDM} signal is amplified and through an \gls{AOM} enters the recirculating loop which consists of a \gls{LSPS}, a 75-km span of \gls{SSMF}, an \gls{EDFA}, an \gls{AOM} and an \gls{OTF} used for gain flattening. The inset of Fig.~\ref{fig:setup}   shows the optical spectrum after 80 circulations, which corresponds to 6000~km of transmission using only \gls{EDFA}-amplification. {Optionally, a hybrid amplification scheme can be used by adding a 750~mW 1445~nm Raman pump in a backward pumping configuration.} The output of the recirculating loop is amplified, filtered by a \gls{WSS} and digitized by a coherent receiver consisting of a \gls{LO}, a 90-degree hybrid, four balanced photo-diodes and an 80 GSa/s \gls{ADC}. 
The offline \gls{DSP} includes front-end impairment correction using blind moment estimation, chromatic dispersion compensation, frequency offset estimation and correction between transmitter and \gls{LO} laser. A widely-linear \gls{MIMO} equalization \cite{daSilvaJLT2016} with \gls{BPS} \cite{PfauJLT2009} inside the update loop is employed to correct for phase noise, error counting and \gls{GMI} evaluation.
In the following sections, we evaluate and discuss the results  for two configurations as EDFA-only amplification (Sec. \ref{sec:exp_EDFA}) and a hybrid of EDFA and Raman (Sec. \ref{sec:exp_Hybrid}).

\subsection{Experimental Results: EDFA-only Amplifier}\label{sec:exp_EDFA}
Fig. \ref{fig:constellation_B2B} and Fig.  \ref{fig:constellation_70spans} show the 2D projections of the received constellations after optical back-to-back and after transmission over 70 spans respectively. Note that the proposed 4D-OS128 modulation  induces nonuniform  probability  distribution  when projected onto 2D, which is similar to probabilistic amplitude shaped 16QAM with very short  blocklength of 4. Instead of using four time slots as PS, the  4D-OS128 shapes the constellation using the two  quadratures  (I/Q)  and  the  two  polarization  states (X/Y) as four dimensions.

\begin{figure}[!tb]
\centering
\subfigure[Average \gls{GMI}  per channel versus transmission distance at  {total} launch power of 9.5~dBm. Inset: Average \gls{GMI}  per channel versus total launch power after 6000~km.]{
	\hspace{-0.5em}
\includegraphics[width=0.49\textwidth]{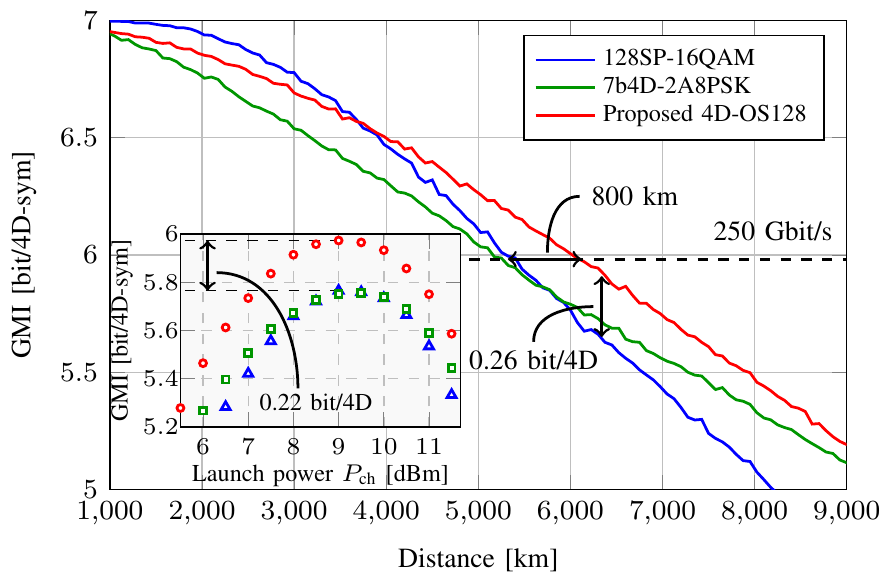}}

\subfigure[BER versus transmission distance at  {total} launch power of 9.5~dBm.]{\includegraphics[width=0.49\textwidth]{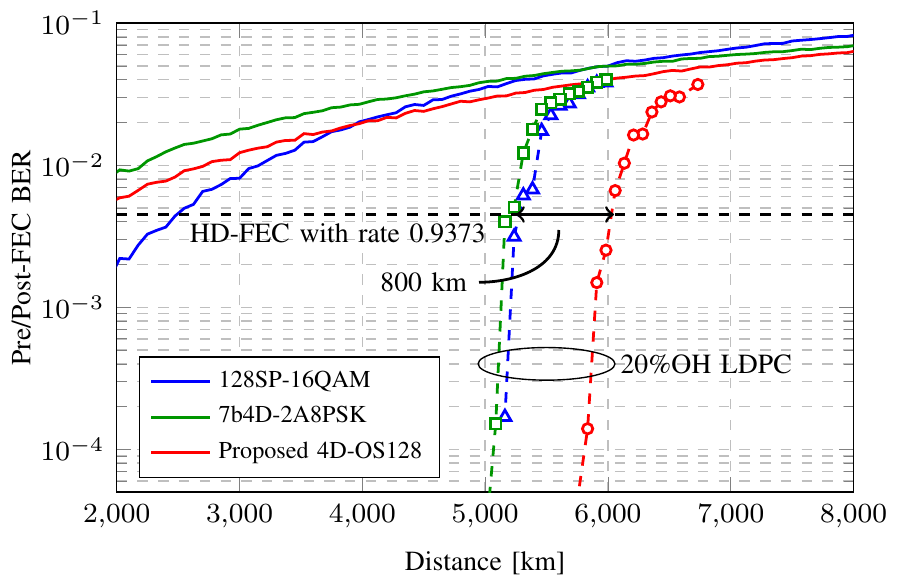}}
   \vspace{-1em}
\caption{Experimental results using EDFA-only amplification.}
    \label{fig:experiment_results}
\vspace{-1em}
\end{figure}

Fig. \ref{fig:experiment_results} (a) shows the GMI as a function of transmission distance for the optimal  {total}  launch power of 9.5 dBm. The GMI in Fig. \ref{fig:experiment_results} (a) should
be interpreted as the reach that an ideal SD-FEC would achieve. 
 For the considered rate (6 bit/4D-sym), 7b4D-2A8PSK offers
approximately  the same reach as 128SP-16QAM around 5320~km. 
The proposed 4D-OS128  reaches  6120~km, which corresponds to a gain of 800~km (15\%). 
This observed reach increase in percentage  is in good agreement with the simulation results of Fig. \ref{fig:comparison_7bitmodulation}.\footnote{Note that the experimental results shows a shorter transmission distance than the simulations in Fig. \ref{fig:comparison_7bitmodulation}. This is due to the  loss from optical components (LSPS, AOM and OTF)  in the recirculating loop.} 
These experimental results confirm the performance
of the novel 4D-OS128 modulation format obtained in
simulation.
Fig. \ref{fig:experiment_results} (a) also shows that  the average \gls{GMI}  per channel resulting in a 0.26~bit/4D increase for 4D-OS128 with respect to 128SP-16QAM after 6340~km transmission.
The GMI vs. launch power for a transmission distance of 6000~km and the three modulation formats under consideration are also shown as inset in Fig. \ref{fig:experiment_results} (a).
At the optimal launch power, 4D-OS128 outperforms  128SP-16QAM and 7b4D-2A8PSK with a gain of  0.22~bits/4D-sym.

BER performance before and after FEC are shown in Fig. \ref{fig:experiment_results} (b).
For the experiment,  low-density
parity-check (LDPC) blocks are constructed, which are then encoded using the DVB-S2 LDPC
code with 20\% overhead and code length $n=64800$.
An outer hard decision forward error correction (HD-FEC) staircase
code with rate 0.9373 \cite{Smith2012}  that corrects bit errors after LDPC decoding is assumed.
The BER threshold is $4.5 \times 10^{-3}$
\cite[Fig. 8]{Smith2012}, which makes 4D-OS128 15\% (+800~km) better in reach (6050~km)  compared to 128SP-16QAM  (5250~km) with the same data rate. 
Moreover, the 15\% reach increase is preserved for the post-FEC gain using an off-the-shelf DVB-S2 LDPC.
In addition to the gains shown versus 128SP-16QAM,  4D-OS128 is also shown to outperform 7b4D-2A8PSK.
The combination of the rate of this staircase code together with a baudrate of 41.79 GBd and a net data rate of 5.58 bits/4D-sym results in a total data rate of just over 233~Gbit/s per channel.

\begin{figure}[!tb]
\vspace{0em}
\centering
\includegraphics[width=0.43\textwidth]{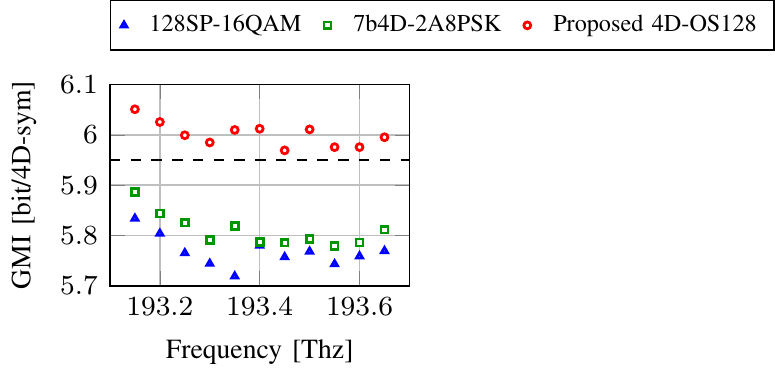}
\hspace{-10.5em}\includegraphics[width=0.225\textwidth]{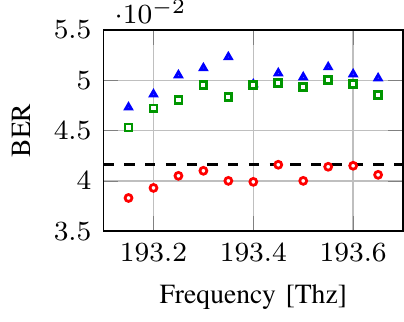}
   \vspace{-0.5em}
\caption{Experimental results using EDFA-only amplification. Per-channel performance versus GMIs (left) and BERs (right) measured for all 11 channels individually after 6000~km showing  GMIs above 5.95 bit/4D \cite{Kojima2017JLT} and BERs below the FEC threshold $4.1\cdot10^{-2}$ \cite{SugiharaOFC13}.}
    \label{fig:experiment_channelsweep_EDFA}
    \vspace{-1em}
\end{figure}

At this optimal launch power, we demonstrate transmission  above the \gls{GMI}  threshold of 5.95~bit/4D-sym (0.85 NGMI) enabling  6000~km error-free  transmission of net 233 Gbit/s per channel after 25.5\% overhead and for all 11 channel in Fig. \ref{fig:experiment_channelsweep_EDFA}. The GMI threshold 5.95~bit/4D-sym (0.85 NGMI) is based on a spatially-coupled type LDPC code\cite{SugiharaOFC13} and the corresponding BER threshold of $4.1\cdot10^{-2}$ is derived in \cite{Kojima2017JLT}.
Note that the rate loss due to the practical FEC after 6000~km transmission is 16 Gbit/s between the net data rate of  233 Gbit/s and the maximum data rate  249 Gbit/s (41.79~GBaud$\times$5.95~bit/4D-sym). Therefore, the relative FEC loss is 6.4\%.

\subsection{Experimental Results: Hybrid  Amplification}\label{sec:exp_Hybrid}
In the hybrid amplification scheme, 
a 750~mW 1445~nm Raman pump is also used  in a backward configuration  as shown in Fig. \ref{fig:setup}.

\begin{figure}[!tb]
\centering
\subfigure[Average \gls{GMI}  per channel versus transmission distance at  {total} launch power of 6.5~dBm. Inset: Average \gls{GMI}  per channel versus total launch power after 9000~km.]{	\hspace{-2em}
\includegraphics[width=0.47\textwidth]{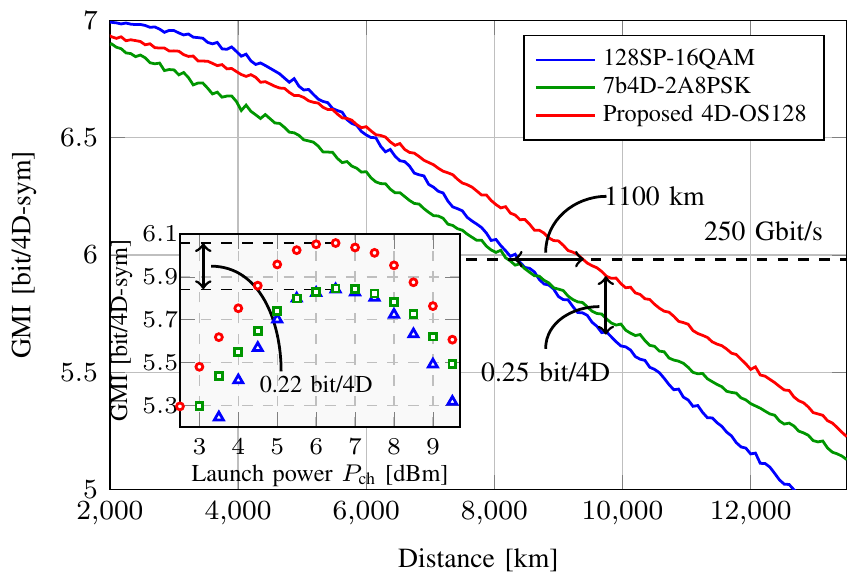}}

\subfigure[BER versus transmission distance at  {total} launch power of 6.5~dBm.]{\includegraphics[width=0.49\textwidth]{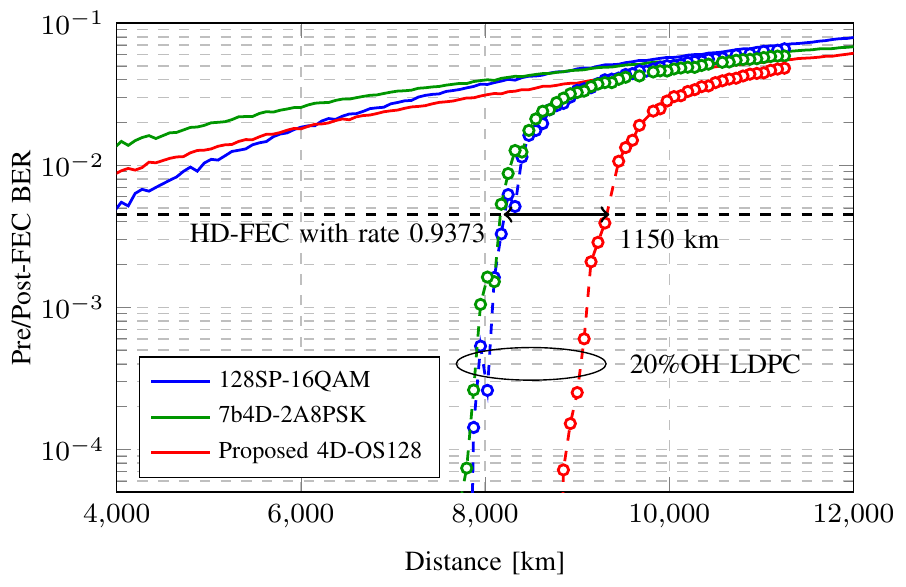}}
   \vspace{-1em}
\caption{Experimental results using Hybrid  amplification.}
    \label{fig:experiment_results_Raman}
    \vspace{-1.2em}
\end{figure}

The  GMI  vs. launch power for a transmission distance of 9000~km and the three modulation formats under consideration are also shown as  inset  in
 Fig. \ref{fig:experiment_results_Raman} (a).
At  the  optimal  launch  power of 6.5~dBm, 
4D-OS128  maximizes the average GMI per channel resulting in a 0.22 bits/4D-sym increase for  with respect to   128SP-16QAM  and  7b4D-2A8PSK.
 Therefore, we use  6.5  dBm as launch  power to evaluate the transmission performance.
 Fig. \ref{fig:experiment_results_Raman} (a). shows  the  GMI  as  a  function  of  transmission distance.
We can observe that  the relative GMI gains of 0.25 bit/4D-sym  and 1100~km (13.5\%) are similar to the EDFA-only case.

\begin{figure}[!tb]
\centering
\includegraphics[width=0.43\textwidth]{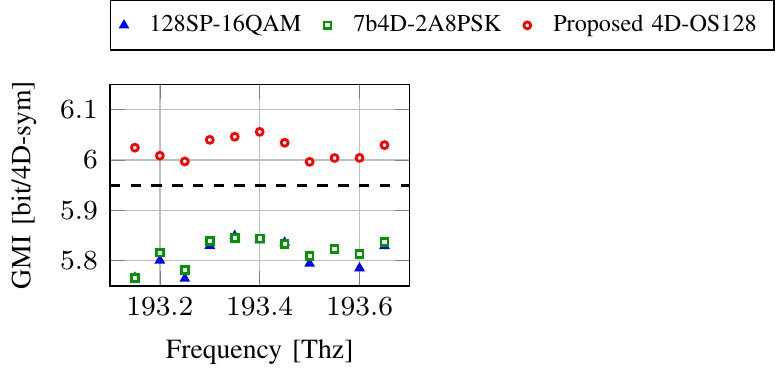}\hspace{-10.05em}\includegraphics[width=0.225\textwidth]{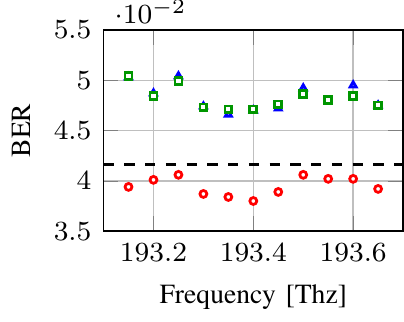}
   \vspace{-0.5em}
\caption{Experimental results using a hybrid of EDFA and Raman amplification. Per-channel performance versus GMIs (left) and BERs (right) measured for all 11 channels individually after 9000~km showing  GMIs above 5.95 bit/4D \cite{Kojima2017JLT} and BERs below the FEC threshold $4.1\cdot10^{-2}$ \cite{SugiharaOFC13}.}
    \label{fig:experiment_channelsweep_hybrid}
       \vspace{-1em}
\end{figure}

BER performance before and after FEC are shown in Fig. \ref{fig:experiment_results_Raman} (b). 
Under the assumption of concatenated LDPC and staircase  code  with  rate  0.9373, 4D-OS128
 shows a post-FEC reach increase of 1150~km (14\%). By measuring all the 11 \gls{WDM} channels’ performance, Fig. \ref{fig:experiment_channelsweep_hybrid} shows all channels above   the  GMI  threshold  of  5.95  bit/4D-sym  (0.85  NGMI) enabling 9000~km  error-free  transmission  of  net 233  Gbit/s per channel after 25.5\% overhead. 
Comparing to the results of the EDFA-only amplification in Sec. \ref{sec:exp_EDFA}, 
utilizing Raman amplifier causes an apparent improvement (50\% reach increase) on the transmission performance compared to using the  EDFA for the 4D-OS128 modulation.

\section{Conclusions}\label{con}
A new modulation format (4D-OS128) with a spectral efficiency of 7 bits/4D-sym was introduced using the concept of orthant symmetry. The format was designed based on the generalized mutual information, and thus, it finds applications to systems with soft-decision forward error correcting codes and bit-wise decoding. The 4D-OS128  format provides sensitivity gains of up to 0.65~dB after LDPC decoding versus   {two well-studied 4D modulation formats: 128SP-16QAM and 7b4D-2A8PSK.} 
 {The numerical results of optical multi-span transmission results  shows significant improvement over GS formats and also a comparable performance versus  PS-QAM with short blocklength for distribution matching. The experimental results confirm the overall superior receiver sensitivity of  4D-OS128 versus previously published GS formats.} Transmission reach extensions of more than 15\% is demonstrated. We believe that the proposed format is a good alternative for future high capacity long haul transmission systems, which provides an intermediate solution between PM-8QAM and PM-16QAM. The design of orthant-symmetric constellations for higher dimensions (e.g., 16 dimensions) and larger constellation sizes (e.g., 256-ary, 512-ary and 1024-ary formats) are left for further investigation.
 
\begin{appendices} 
\section{Coordinates and binary labeling for 4D-OS128}

Table \ref{tab:4D_128_XL} lists the coordinates of the constellation points and the bit-to-symbol mapping of 4D-OS128  {for the target SNR of 9.5~dB}. The constellation is assumed to be  normalized to $E_s=2$, i.e., to unit energy per polarization.

\input{4D128_XL.tex}

\end{appendices}

\section*{Acknowledgement}
Fraunhofer HHI and ID Photonics are gratefully acknowledged for providing their Optical-Multi-Format Transmitter.

\bibliographystyle{IEEEtran}
\bibliography{references_4D64PRS,references}
\balance
\end{document}

%% file: myFigures.tex
\usetikzlibrary{plotmarks,matrix,chains,scopes,fit,calc,shapes,positioning,decorations,intersections,arrows,backgrounds,shadows}

\newlength\FigureHeight
\newlength\FigureWidth


%% file: myStyles.tex
\definecolor{myDarkGreen}{rgb}{0.00000,0.58824,0.00000}%
\definecolor{uniform}{rgb}{0.00000,0.58824,0.00000}%
\definecolor{AWGNreference}{rgb}{0.00000,0.58824,0.00000}%

%% file: myMacros.tex
\DeclareMathOperator*{\argmax}{argmax}

\newcounter{lemma}

\newtheorem{exampleplain}{Example}

\newenvironment{example}{\begin{exampleplain}}{~\hfill$\vartriangle$\end{exampleplain}}
\newtheorem{definitionplain}{Definition}
\newenvironment{definition}{\begin{definitionplain}}{~\hfill$\square$\end{definitionplain}}

\newcommand{\mcIkb}{\mathcal{I}_{k}^{b}}

\newcommand{\set}[1]{\{#1\}}

\newcommand{\ld}{\ldots}

\newcommand{\bC}{\boldsymbol{C}}\newcommand{\bc}{\boldsymbol{c}}
\newcommand{\bb}{\boldsymbol{b}}
\newcommand{\bX}{\boldsymbol{X}}\newcommand{\bx}{\boldsymbol{x}}

\newcommand{\bY}{\boldsymbol{Y}}\newcommand{\by}{\boldsymbol{y}}

\newcommand{\bS}{\boldsymbol{S}}\newcommand{\bs}{\boldsymbol{s}}

\newcommand{\un}[1]{\underline{#1}}

\newcommand{\tnr}[1]{{\textnormal{#1}}}



%% file: myPackages.tex
\usepackage[utf8]{inputenc}

\usepackage{amsmath,amssymb}

\usepackage{xcolor}
\usepackage{graphicx}
\usepackage{tikz}
\usepackage{pgfplots}
\usepackage{float}
\usepackage{balance}

\usepackage{xspace}
\usepackage[normalem]{ulem}

\usepackage{dsfont,empheq}
\usepackage{multicol,multirow}

\usepackage{cite}
\usepackage{gensymb}

\usepackage{booktabs, cellspace, hhline}

%% file: acronyms.tex
\newacronym{CUT}{CUT}{channel under test}

\newacronym{KK}{KK}{Kramers-Kronig}
\newacronym{CSPR}{CSPR}{carrier-to-signal power ratio}
\newacronym{KKRX}{KKRx}{Kramers-Kronig Receiver}
\newacronym{SSBI}{SSBI}{signal-signal beat interference}

\newacronym{GS}{GS}{geometric shaping}
\newacronym{PS}{PS}{probabilistic shaping}
\newacronym{DSP}{DSP}{digital signal processing}
\newacronym{MIMO}{MIMO}{multiple-input multiple-output}
\newacronym{TDE}{TDE}{time domain equalizer}
\newacronym{FDE}{FDE}{frequency domain equalizer}
\newacronym{LMS}{LMS}{least means square}
\newacronym{DDLMS}{DD-LMS}{decision directed least means square}
\newacronym{FFE}{FFE}{feed-forward equalizer}
\newacronym{FBE}{FBE}{feedback equalizer}
\newacronym{BPS}{BPS}{blind phase search}

\newacronym{SMF}{SMF}{single-mode fiber}
\newacronym[plural=SSMFs]{SSMF}{SSMF}{standard single-mode fiber}
\newacronym[plural=FMFs]{FMF}{FMF}{few-mode fiber}
\newacronym{FMF12}{FMF12}{12 mode FMF}
\newacronym{MMF}{MMF}{multi-mode fiber}
\newacronym{SI}{SI}{step index}
\newacronym{GI}{GI}{graded index}
\newacronym{DCF}{DCF}{dispersion compensated fiber}

\newacronym{SDM}{SDM}{space division multiplexing}
\newacronym{MDM}{MDM}{mode division multiplexed}
\newacronym{WDM}{WDM}{wavelength division multiplexing}
\newacronym{DWDM}{DWDM}{dense wavelength division multiplexing}
\newacronym{LP}{LP}{linear polarized}
\newacronym[plural=MMUXs,firstplural=mode multiplexers]{MMUX}{MMUX}{mode multiplexer}
\newacronym{PL}{PL}{photonic lantern}
\newacronym{3DWG}{3DWG}{3D-waveguide}
\newacronym{MDL}{MDL}{mode dependent loss}
\newacronym{DGD}{DGD}{differential group delay}
\newacronym{DMGD}{DMGD}{differential mode group delay}
\newacronym{QSM}{QSM}{quasi-single-mode}
\newacronym{GIMMF}{GI-MMF}{graded-index multi-mode fiber}

\newacronym{SSB}{SSB}{single side band}
\newacronym{QPSK}{QPSK}{quadrature phase shift keying}
\newacronym{QAM}{QAM}{quadrature amplitude modulation}
\newacronym{RRC}{RRC}{root-raised-cosine}
\newacronym{4D-64PRS}{4D-64PRS}{
four-dimensional 64-ary polarization-ring-switching}
\newacronym{DP}{DP}{dual-polarization}
\newacronym[\glslongpluralkey=states-of-polarization]{SOP}{SOP}{state-of-polarization}
\newacronym{PM}{PM}{polarization-multiplexed}

\newacronym{ECL}{ECL}{external cavity laser}
\newacronym{CW}{CW}{continuous wave}
\newacronym[plural=DFBs]{DFB}{DFB}{distributed feedback laser}

\newacronym[plural=DACs]{DAC}{DAC}{digital-to-analog converter}
\newacronym{ADC}{ADC}{analog-to-digital converter}
\newacronym{PRBS}{PRBS}{pseudo-random bit sequence}
\newacronym{LO}{LO}{local oscillator}
\newacronym{EDFA}{EDFA}{erbium-doped fiber amplifier}
\newacronym{MZM}{MZM}{Mach-Zehnder modulator}
\newacronym{DP-MZM}{DP-MZM}{dual-polarization Mach-Zehnder modulator}
\newacronym{ChUT}{ChUT}{channel under test}
\newacronym{WSS}{WSS}{wavelength selective switch}
\newacronym[plural=VOAs]{VOA}{VOA}{variable optical attenuator}
\newacronym[plural=PDCRXs]{PDCRX}{PDCRX}{polarization diverse coherent receiver}
\newacronym{DSO}{DSO}{digital storage oscilloscope}
\newacronym{ASE}{ASE}{amplified spontaneous emission}
\newacronym{PBS}{PBS}{polarization beam splitter}
\newacronym{PD}{PD}{photodiode}
\newacronym{AOM}{AOM}{acousto-optic modulator}
\newacronym{BPD}{BPD}{balanced photo-diode}
\newacronym{OMFT}{OMFT}{optical multi-format transmitter}
\newacronym{DPIQ}{DP-IQM}{dual-polarization IQ-modulator}
\newacronym{ABC}{ABC}{automatic bias controller}
\newacronym{OTF}{OTF}{optical tunable filter}
\newacronym{LSPS}{LSPS}{loop-synchronous polarization scrambler}
\newacronym{OSA}{OSA}{optical spectrum analyzer}

\newacronym{OSNR}{OSNR}{optical signal to noise ratio}
\newacronym{BER}{BER}{bit error rate}
\newacronym{IL}{IL}{insertion loss}

\newacronym{SDFEC}{SD-FEC}{soft-decision forward error correction}
\newacronym{HDFEC}{HD-FEC}{hard-decision forward error correction}
\newacronym{FEC}{FEC}{forward error correction}
\newacronym{LDPC}{LDPC}{low-density parity-check code}

\newacronym{AIR}{AIR}{achievable information rate}
\newacronym{AIRs}{AIRs}{achievable information rates}
\newacronym{AR}{AR}{achievable rates}
\newacronym{MI}{MI}{mutual information}
\newacronym{GMI}{GMI}{generalized mutual information}
\newacronym{BICM}{BICM}{bit-interleaved coded modulation}

\newacronym{CM}{CM}{coded modulation}

\newacronym{OVNA}{OVNA}{optical vector network analyzer}
\newacronym{NIR}{NIR}{near infrared}
\newacronym{CD}{CD}{chromatic dispersion}
\newacronym{OTDR}{OTDR}{optical time domain reflectometry}
\newacronym{OFDR}{OFDR}{optical frequency domain reflectometry}
\newacronym{GPU}{GPU}{graphics processing unit}
\newacronym{SVD}{SVD}{singular value decomposition}
\newacronym{WGN}{WGN}{white Gaussian noise}
\newacronym{AWGN}{AWGN}{additive white Gaussian noise}
\newacronym{PDL}{PDL}{polarization dependent loss}
\newacronym{SPS}{sps}{samples-per-symbol}
\newacronym{SE}{SE}{spectral efficiency}

%% file: 4D128_XL.tex





\begin{table}[!htbp]
\caption{Coordinates and binary labeling of the proposed 4D-OS128 format for the target SNR of 9.5~dB. The  coordinates are rounded to four decimal points: $(t_1,t_2,t_3,t_4,t_5)=(0.2875, 0.3834,0.4730,1.1501,1.2460)$}
\label{tab:4D_128_XL}
\renewcommand{\arraystretch}{1.05}
\centering {\footnotesize
\begin{tabular}{|@{}|@{\hskip 1ex}c@{\hskip 1ex}@{\hskip 1ex}c@{\hskip 1ex}|@{\hskip 1ex}c@{\hskip 1ex}|@{\hskip 1ex}c@{\hskip 1ex}|@{}}
\hline

\hline
 Coordinates & Labeling & Coordinates & Labeling\\ 
\hline 
$(+t_3,+t_3,+t_1,+t_1)$ & 0000011&   
$(+t_2,+t_5,+t_3,+t_3)$ & 0000001\\ \hline  
$(-t_3,+t_3,+t_1,+t_1)$ & 1000011&   
$(-t_2,+t_5,+t_3,+t_3)$ & 1000001\\ \hline  
$(-t_3,-t_3,+t_1,+t_1)$ & 1100011&   
$(-t_2,-t_5,+t_3,+t_3)$ & 1100001\\ \hline  
$(+t_3,-t_3,+t_1,+t_1)$ & 0100011&   
$(+t_2,-t_5,+t_3,+t_3)$ & 0100001\\ \hline  
$(+t_1,+t_1,+t_3,+t_3)$ & 0000111&   
$(+t_3,+t_3,+t_5,+t_2)$ & 0000101\\ \hline  
$(-t_1,+t_1,+t_3,+t_3)$ & 1000111&   
$(-t_3,+t_3,+t_5,+t_2)$ & 1000101\\ \hline  
$(-t_1,-t_1,+t_3,+t_3)$ & 1100111&   
$(-t_3,-t_3,+t_5,+t_2)$ & 1100101\\ \hline  
$(+t_1,-t_1,+t_3,+t_3)$ & 0100111&   
$(+t_3,-t_3,+t_5,+t_2)$ & 0100101\\ \hline  
$(+t_3,+t_3,+t_2,+t_5)$ & 0000110&   
$(+t_3,+t_3,+t_4,+t_4)$ & 0000100\\ \hline  
$(-t_3,+t_3,+t_2,+t_5)$ & 1000110&   
$(-t_3,+t_3,+t_4,+t_4)$ & 1000100\\ \hline  
$(-t_3,-t_3,+t_2,+t_5)$ & 1100110&   
$(-t_3,-t_3,+t_4,+t_4)$ & 1100100\\ \hline  
$(+t_3,-t_3,+t_2,+t_5)$ & 0100110&   
$(+t_3,-t_3,+t_4,+t_4)$ & 0100100\\ \hline  
$(+t_5,+t_2,+t_3,+t_3)$ & 0000010&   
$(+t_4,+t_4,+t_3,+t_3)$ & 0000000\\ \hline  
$(-t_5,+t_2,+t_3,+t_3)$ & 1000010&   
$(-t_4,+t_4,+t_3,+t_3)$ & 1000000\\ \hline  
$(-t_5,-t_2,+t_3,+t_3)$ & 1100010&   
$(-t_4,-t_4,+t_3,+t_3)$ & 1100000\\ \hline  
$(+t_5,-t_2,+t_3,+t_3)$ & 0100010&   
$(+t_4,-t_4,+t_3,+t_3)$ & 0100000\\ \hline  
$(+t_3,+t_3,-t_1,+t_1)$ & 0010011&   
$(+t_2,+t_5,-t_3,+t_3)$ & 0010001\\ \hline  
$(-t_3,+t_3,-t_1,+t_1)$ & 1010011&   
$(-t_2,+t_5,-t_3,+t_3)$ & 1010001\\ \hline  
$(-t_3,-t_3,-t_1,+t_1)$ & 1110011&   
$(-t_2,-t_5,-t_3,+t_3)$ & 1110001\\ \hline  
$(+t_3,-t_3,-t_1,+t_1)$ & 0110011&   
$(+t_2,-t_5,-t_3,+t_3)$ & 0110001\\ \hline  
$(+t_1,+t_1,-t_3,+t_3)$ & 0010111&   
$(+t_3,+t_3,-t_5,+t_2)$ & 0010101\\ \hline  
$(-t_1,+t_1,-t_3,+t_3)$ & 1010111&   
$(-t_3,+t_3,-t_5,+t_2)$ & 1010101\\ \hline  
$(-t_1,-t_1,-t_3,+t_3)$ & 1110111&   
$(-t_3,-t_3,-t_5,+t_2)$ & 1110101\\ \hline  
$(+t_1,-t_1,-t_3,+t_3)$ & 0110111&   
$(+t_3,-t_3,-t_5,+t_2)$ & 0110101\\ \hline  
$(+t_3,+t_3,-t_2,+t_5)$ & 0010110&   
$(+t_3,+t_3,-t_4,+t_4)$ & 0010100\\ \hline  
$(-t_3,+t_3,-t_2,+t_5)$ & 1010110&   
$(-t_3,+t_3,-t_4,+t_4)$ & 1010100\\ \hline  
$(-t_3,-t_3,-t_2,+t_5)$ & 1110110&   
$(-t_3,-t_3,-t_4,+t_4)$ & 1110100\\ \hline  
$(+t_3,-t_3,-t_2,+t_5)$ & 0110110&   
$(+t_3,-t_3,-t_4,+t_4)$ & 0110100\\ \hline  
$(+t_5,+t_2,-t_3,+t_3)$ & 0010010&   
$(+t_4,+t_4,-t_3,+t_3)$ & 0010000\\ \hline  
$(-t_5,+t_2,-t_3,+t_3)$ & 1010010&   
$(-t_4,+t_4,-t_3,+t_3)$ & 1010000\\ \hline  
$(-t_5,-t_2,-t_3,+t_3)$ & 1110010&   
$(-t_4,-t_4,-t_3,+t_3)$ & 1110000\\ \hline  
$(+t_5,-t_2,-t_3,+t_3)$ & 0110010&   
$(+t_4,-t_4,-t_3,+t_3)$ & 0110000\\ \hline  
$(+t_3,+t_3,-t_1,-t_1)$ & 0011011&   
$(+t_2,+t_5,-t_3,-t_3)$ & 0011001\\ \hline  
$(-t_3,+t_3,-t_1,-t_1)$ & 1011011&   
$(-t_2,+t_5,-t_3,-t_3)$ & 1011001\\ \hline  
$(-t_3,-t_3,-t_1,-t_1)$ & 1111011&   
$(-t_2,-t_5,-t_3,-t_3)$ & 1111001\\ \hline  
$(+t_3,-t_3,-t_1,-t_1)$ & 0111011&   
$(+t_2,-t_5,-t_3,-t_3)$ & 0111001\\ \hline  
$(+t_1,+t_1,-t_3,-t_3)$ & 0011111&   
$(+t_3,+t_3,-t_5,-t_2)$ & 0011101\\ \hline  
$(-t_1,+t_1,-t_3,-t_3)$ & 1011111&   
$(-t_3,+t_3,-t_5,-t_2)$ & 1011101\\ \hline  
$(-t_1,-t_1,-t_3,-t_3)$ & 1111111&   
$(-t_3,-t_3,-t_5,-t_2)$ & 1111101\\ \hline  
$(+t_1,-t_1,-t_3,-t_3)$ & 0111111&   
$(+t_3,-t_3,-t_5,-t_2)$ & 0111101\\ \hline  
$(+t_3,+t_3,-t_2,-t_5)$ & 0011110&   
$(+t_3,+t_3,-t_4,-t_4)$ & 0011100\\ \hline  
$(-t_3,+t_3,-t_2,-t_5)$ & 1011110&   
$(-t_3,+t_3,-t_4,-t_4)$ & 1011100\\ \hline  
$(-t_3,-t_3,-t_2,-t_5)$ & 1111110&   
$(-t_3,-t_3,-t_4,-t_4)$ & 1111100\\ \hline  
$(+t_3,-t_3,-t_2,-t_5)$ & 0111110&   
$(+t_3,-t_3,-t_4,-t_4)$ & 0111100\\ \hline  
$(+t_5,+t_2,-t_3,-t_3)$ & 0011010&   
$(+t_4,+t_4,-t_3,-t_3)$ & 0011000\\ \hline  
$(-t_5,+t_2,-t_3,-t_3)$ & 1011010&   
$(-t_4,+t_4,-t_3,-t_3)$ & 1011000\\ \hline  
$(-t_5,-t_2,-t_3,-t_3)$ & 1111010&   
$(-t_4,-t_4,-t_3,-t_3)$ & 1111000\\ \hline  
$(+t_5,-t_2,-t_3,-t_3)$ & 0111010&   
$(+t_4,-t_4,-t_3,-t_3)$ & 0111000\\ \hline  
$(+t_3,+t_3,+t_1,-t_1)$ & 0001011&   
$(+t_2,+t_5,+t_3,-t_3)$ & 0001001\\ \hline  
$(-t_3,+t_3,+t_1,-t_1)$ & 1001011&   
$(-t_2,+t_5,+t_3,-t_3)$ & 1001001\\ \hline  
$(-t_3,-t_3,+t_1,-t_1)$ & 1101011&   
$(-t_2,-t_5,+t_3,-t_3)$ & 1101001\\ \hline  
$(+t_3,-t_3,+t_1,-t_1)$ & 0101011&   
$(+t_2,-t_5,+t_3,-t_3)$ & 0101001\\ \hline  
$(+t_1,+t_1,+t_3,-t_3)$ & 0001111&   
$(+t_3,+t_3,+t_5,-t_2)$ & 0001101\\ \hline  
$(-t_1,+t_1,+t_3,-t_3)$ & 1001111&   
$(-t_3,+t_3,+t_5,-t_2)$ & 1001101\\ \hline  
$(-t_1,-t_1,+t_3,-t_3)$ & 1101111&   
$(-t_3,-t_3,+t_5,-t_2)$ & 1101101\\ \hline  
$(+t_1,-t_1,+t_3,-t_3)$ & 0101111&   
$(+t_3,-t_3,+t_5,-t_2)$ & 0101101\\ \hline  
$(+t_3,+t_3,+t_2,-t_5)$ & 0001110&   
$(+t_3,+t_3,+t_4,-t_4)$ & 0001100\\ \hline  
$(-t_3,+t_3,+t_2,-t_5)$ & 1001110&   
$(-t_3,+t_3,+t_4,-t_4)$ & 1001100\\ \hline  
$(-t_3,-t_3,+t_2,-t_5)$ & 1101110&   
$(-t_3,-t_3,+t_4,-t_4)$ & 1101100\\ \hline  
$(+t_3,-t_3,+t_2,-t_5)$ & 0101110&   
$(+t_3,-t_3,+t_4,-t_4)$ & 0101100\\ \hline  
$(+t_5,+t_2,+t_3,-t_3)$ & 0001010&   
$(+t_4,+t_4,+t_3,-t_3)$ & 0001000\\ \hline  
$(-t_5,+t_2,+t_3,-t_3)$ & 1001010&   
$(-t_4,+t_4,+t_3,-t_3)$ & 1001000\\ \hline  
$(-t_5,-t_2,+t_3,-t_3)$ & 1101010&   
$(-t_4,-t_4,+t_3,-t_3)$ & 1101000\\ \hline  
$(+t_5,-t_2,+t_3,-t_3)$ & 0101010&   
$(+t_4,-t_4,+t_3,-t_3)$ & 0101000\\ \hline  

\hline 
\end{tabular}}
\end{table}